\documentclass[useAMS,usenatbib,usegraphicx]{mn2e}
\usepackage{graphicx}
\usepackage{color}

\title[Messier 5: I. Period changes]{Long-term photometric monitoring of Messier 5 variables:\\ I. Period changes of RR Lyrae stars}

\author[Szeidl et al.]{B. Szeidl$^{1}$, Zs. Hurta$^{1}$, J. Jurcsik$^{1}$, C. Clement$^{2}$ and M. Lovas$^{1}$ \\
$^{1}$Konkoly Observatory of the Hungarian Academy of Sciences, H-1525 Budapest PO Box 67, Hungary\\
$^{2}$Dept. of Astronomy, University Toronto, Ontario, M5S3H8, Canada}
\begin{document}

\date{Accepted 2010 Sept 29  Received 2010 Sept 27; in original form 2010 Aug 3}

\pagerange{\pageref{firstpage}--\pageref{lastpage}} \pubyear{2010}

\maketitle

\label{firstpage}

\begin{abstract}
The period changes of 86 M5 RR Lyrae stars have been investigated on a one-hundred-year time base. The published observations have been supplemented by archival Asiago, Konkoly and Las Campanas photographic observations obtained between 1952 and 1993. About two thirds of the $O-C$ diagrams could be fitted by a straight line or a parabola. 21 RR Lyrae stars have increasing, 18 decreasing and 16 constant period. The mean rates of period change of these variables are:
 $\langle \beta \rangle = \langle \dot{P}\rangle  = -0.006\pm0.162 \, \mathrm{d {Myr}^{-1}}, \langle \alpha \rangle =\langle P^{-1}\dot{P}\rangle  = -0.021 \pm0.308 \, \mathrm{Myr}^{-1}$. Ten RR Lyrae stars show fast period decrease with $\dot{P} < - 0.10 \,  \mathrm{d{Myr}^{-1}}$.
At least some of these variables may be in the pre-zero-age horizontal-branch (ZAHB) evolutionary stage.
The variables on the long-period sequence of the period-amplitude diagram are brighter than the other RR Lyrae stars of M5 and are in an advanced evolutionary stage moving off from the HB redward. More than one third of the M5 RR Lyrae stars investigated have irregular period change.  The irregular period behaviour is relatively more frequent among the RRc (RR1) stars (50 per cent) than among the RRab (RR0) stars (34 per cent). A strict relationship has been found between the irregular period change and the Blazhko effect of M5 RRab stars. This fact indicates a common origin for these phenomena. It is remarkable that, if the RRab stars showing  Blazhko effect are omitted from the sample, the mean rates of the period change have small positive values ($\langle \beta \rangle=0.012 \pm0.147 \, \mathrm{d {Myr}^{-1}}$, $\langle \alpha \rangle=0.013 \pm 0.279 \, \mathrm{Myr}^{-1}$), in excellent agreement with HB evolutionary model predictions. 

\end{abstract}

\section{Introduction}

More than ninety years ago \citet{ed18} had already realized that the period changes of a Cepheid-like pulsator would give information on the changes of the physics of the star's interior during its evolution. The potential of measuring the rate of stellar evolution seemed to be fulfilled when \citet{ma38} discovered that the periods of the RRab stars in omega Centauri were predominantly increasing. The subsequent observations of RR Lyrae stars in other globular clusters, however, failed to demonstrate the direct connection between the detected period changes and horizontal-branch (HB) evolution. The observed large period decreases, the random and/or abrupt period changes could not be reconciled with steady evolutionary effects. It has been, however, argued that evolutionary period changes must be present in individual stars and must be observable, at least on long time-scales.Globular clusters with dozens, hundreds of RR Lyrae stars are ideal targets for such studies, the mean value of the period-change rates of a large sample of RR Lyrae stars of a globular cluster must show the trend of their evolution. Indeed, \citet{le91} proved that the mean period-change rates observed in globular clusters were consistent with synthetic HB models, if the HB-types of the clusters were also considered. 

The generally adopted procedure for investigating the period changes of variable stars is the construction and interpretation of their $O-C$ diagrams. This plots the difference between the observed times ($O$) of a particular phase, usually the maximum or the mid-point on the ascending branch of the light curve and the predicted time of the same phase ($C$), calculated according to an accepted ephemeris. However, using today's computational techniques, the $O-C$ data are often calculated as the time shift between the entire observed light curve and the normal light curve, rather than from one specific phase. This method gives more stable results. If the $O-C$ diagram is a linear or quadratic function of time, then the period is constant or continuously changing at a constant rate.

In order to explain the complex nature of some $O-C$ diagrams, i.e. the irregular behaviour of the period variations that masks the evolutionary changes, several suggestions have been put forward in the past. The accumulation of random variations (some sort of noise) in the period may show different drifts in the $O-C$ diagram as a typical realization of a random walk, even if the period noise has a Gaussian distribution (see e.g. \citealt{st34}; \citealt{bd65}; \citealt{lk93}). The appearance of this effect on the $O-C$ diagram, however, strongly depends on the parameters of the probability density function of the period noise, and it does not reflect real period changes. \citet{sr79} suggested a mechanism for the occurrence of irregularities in the pulsation period. According to their theory, random mixing events through the semiconvective zone cause the composition profile of the star's core to be altered. This produces changes in its hydrostatic structure, resulting in abrupt and continuous period changes of both signs. \citet{st80} attributed the irregular period changes to hydromagnetic effects taking place in the atmosphere of the pulsating star. \citet{la74} and \citet{k94} modelled the effect of mass loss on the period changes of RR Lyrae stars, while \citet{co71a} noticed that binarity might lead to long-term oscillation in the star's observed pulsation period.

Messier 5 (NGC 5904) is one of the globular clusters that are richest in RR Lyrae stars. The period changes of the cluster's RR Lyrae stars were already investigated by \citet{cs69}, \citet{kk71}, \citet{st91} and \citet{re96} using baselines  ranging from seventy to one hundred years. In these previous studies, no photometric data were available for nearly two decades between 1970 and 1990, and the most recent period-change study \citep{re96} concerned only 30 variables. Therefore, with the addition of archival data, the reexamination of the period changes of a larger sample of variables is desirable.

Now the observational baseline extends over a century, and the previously published observations, combined with the unpublished ones from the Las Campanas and Konkoly Observatories, render almost continuous coverage for the last fifty years of the 20th century. This provides the motivation for the present study of period changes of RR Lyrae stars in the globular cluster M5.

\section{Data and method of investigation}

\subsection{Previous photometric studies}

The first investigation of the variable stars in M5 was carried out by \citet{ba17}. He made use of 111 plates obtained during the years 1895 to 1908 with the 13-inch Boyden, the 11-inch Draper and the 24-inch Bruce telescopes mostly at Arequipa, Peru with exposure times ranging from 20 to 100 min. This collection was supplemented with 12 photographs taken at the Mount Wilson 60-inch reflector with 10--14 min exposure times in 1912. Bailey published brightness estimates for about eighty RR Lyrae stars, and these observations provide a firm starting point for the study of period changes of the RR Lyrae stars in M5.

\citet{sh27} studied the variables on the basis of a collection of 59 plates mostly with double exposures (in all 113 exposures) taken with the Mount Wilson 60-inch telescope on eight different nights in 1917. Shapley did not publish his measurements, and likely they have gone astray. Therefore, the plates were measured anew, and magnitudes of 62 RR Lyrae stars were published by \citet{cs69} and \citet{co71b}.

A detailed analysis of the variables in M5 was made by \citet{oo41}. He investigated the period changes of the cluster variables, and noted that the light curve of some stars showed RW Dra-type variation (now called Blazhko effect). The observations consisted of 74 and 7 plates taken with the 60-inch reflector of the Mount Wilson observatory in the years 1934 and 1935, respectively. The brightness of 92 RR Lyrae stars were measured or estimated on this material.

Comprehensive studies of the period changes of RR Lyraes in M5 have been carried out by \citet{cs69} and \citet{kk71}, independently of each another.

Coutts \& Sawyer Hogg published magnitudes of 66 RR Lyrae variables of M5 measured on 157 plates taken with the 74-inch telescope of the David Dunlap Observatory between 1936 and 1966.

Kukarkin \& Kukarkina's material (181 plates) was collected with the 40-cm astrograph at Crimea in the years 1952--1968. Due to the severe crowding effect near the centre of the cluster and the close companions in some cases, the brightness of only 51 RR Lyrae variables could be estimated.

Since the mid-eighties of the last century, accurate CCD observations have been obtained for the RR Lyrae variables of M5.

\citet{cg87} observed six variables in $B$ and $I$ on six nights in 1986. \citet{st91} published $B$ and $V$ magnitudes for eleven RR Lyrae stars obtained during four nights in 1987. \citet{br96} and \citet{cm92} investigated the M5 variables in 1989. Cohen \& Matthews made $V,I$ photometry of eight stars on six nights, while Brocato et al. gave measurements in $B,V$ colours for fifteen variables on four nights.

\citet{re96} and \citet{ka00} published comprehensive CCD photometry of the variable stars in M5. Reid obtained his $V,I$ observations during ten nights in 1991 and 1992 and investigated 49 RR Lyrae stars. Kaluzny et al. observed 65 RR Lyrae stars of M5 in $V$ colour. Their observations were made during 12 nights in 1997. We note here that the star identified by Kaluzny as V92 is in fact V17.

\subsection{Observations published in this paper}

In this paper we publish the photographic observations obtained with the University of Toronto 61-cm telescope at the Las Campanas Observatory of the Carnegie Institution of Washington between 1972 and 1991, with the 60-cm reflector of the Konkoly Observatory at Budapest during the years 1952--1963, with the 1-m RCC telescope of the Konkoly Observatory at Piszk\'estet\H{o} Mountain Station from 1976 to 1993 and with the 1.2-m reflector of the Asiago Observatory of University Padova in 1971.

The Asiago, Konkoly and Las Campanas photographic magnitudes of the M5 variables are given in Table~\ref{mag}, available in its full content as Supplementary Material in the electronic edition of this article.

The Las Campanas plates had 103aO emulsion and were exposed through a GG 385 filter, so the magnitudes are $B$ magnitudes. A total of 508 plates was obtained, and they were all measured on a Cuffey iris photometer \citep{c56,c61}. The plates from 1972 to 1980 were measured by summer students, and the data were used, but not published, in the study of the Fourier parameter $\varphi_{31}$ by \cite{cj92}. The 1981 to 1991 plates were measured later by C. Clement. Since the plates were measured at different times by different people, there might be systematic magnitude shifts from year to year, particularly for stars that were subject to crowding. The 1981--1991 magnitudes should be on a consistent system.
However, even then, for crowded stars, it was not always possible to be consistent because of differences in the seeing conditions. Altogether 73 RR Lyrae stars were measured. 

The photographic observations at Budapest with the 60-cm telescope were exposed on Guilleminot Superfulgur or Agfa Astro Spezial plates. Since the sensitivity of the emulsions is similar, the measurements or estimates should be on a consistent system. On 292 plates the brightness of 39 RR Lyrae stars was measured. The magnitudes of a further 35 variables that were close to the crowded central region or had close companion(s) were estimated by M. Lovas.

The observations with the 1-m telescope have been made on 103aO plates. In the years 1976, 1988 and 1992--93 altogether 94 103aO plates were exposed without filters. In the other years 104 photographs were obtained on 103aO plates through a 2mm GG13 filter. In all, 93 variables were measured on these plates. We found no serious systematic difference between the filtered and unfiltered blue magnitudes that exceeded the measurement errors.

The measuring procedure of the photographic plates obtained with the 60-cm and 1-m telescopes was the following. The plates were digitized on a Umax PowerLook 3000 flatbed-transparency scanner with 3.13  and 1.29 pixel-per-arcsec resolution on the 1-m and 60-cm photographs, respectively. Digital aperture photometry was applied to the images using standard IRAF\footnote{{\sc IRAF} is distributed by the National Optical Astronomy Observatories, which are operated by the Association of Universities for Research in Astronomy, Inc., under cooperative agreement with the National Science Foundation.} packages. In this way, the photographic densities of the variables and comparison stars in M5 have been determined.

The brightness of variables on the Asiago plates that were obtained in bad seeing conditions have been estimated by M. Lovas. These plates had IIaO emulsion and were exposed without filter. In all 67 variables were estimated on 35 plates. 

In the course of measuring and estimating the brightness of the variables, the comparison stars were chosen from Arp's  photometry \citep{ar55,ar62}. His photographic ($m_{pg}$) comparison sequence was applied to the Budapest and Asiago plates. The Las Campanas and Piszk\'estet\H{o} observations are on the $B$ magnitude scale. No systematic difference exceeding the 0.02-mag observational error has been found between the filtered and unfiltered Piszk\'estet\H{o} blue observations.

These observations supplemented with the published ones provide an almost continuous coverage for the second half of the last century.

\begin{table}
\caption{Photographic magnitudes of M5 variables. The complete table is given in the electronic version of the article as supporting information (Table S1). }
 \label{mag}
  \begin{tabular}{lccc}
  \hline
star& HJD-2400000 &$B_{pg}$  mag & source$^{a}$\\
 \hline
V1 & 34122.460& 15.41 &  B24\\
.  & ....     & ....  &  ...\\
V1 & 40985.589& 15.80 &  A48\\
.  & ....     &....  &  ...\\
V1 & 41447.629& 15.26 &  L24\\
.  & ....     & ....  &  ...\\
V1 & 42598.453& 15.65 &  P40\\
.  & ....     & ....  &  ...\\
\hline
\multicolumn{4}{l}{\footnotesize{$^{a}$ B24: Budapest 60cm telescope,}}\\
\multicolumn{3}{l}{\footnotesize{ \,\,  A48: Asiago 1.2m telescope,} }\\
\multicolumn{3}{l}{\footnotesize{ \,\,   L24: Las Campanas 61cm telescope,}} \\
\multicolumn{3}{l}{\footnotesize{ \,\,   P40: Piszk\'estet\H{o} 1m telescope.} }\\
\hline
\end{tabular}
\end{table}

\subsection{The method}

\begin{table*}
\begin{minipage}{\textwidth}
\caption{Data.\label{data}}
\begin{tabular}{ccccccccccccccc}
\hline
Star & B17 & S27 & O41 & C69 & B24 & K71 & A48 & L24 & P40 & C87 & S91 & C92 & R96 & K00 \\
\hline
V1 & + & + & + & + & + & + & + & + & + & - & - & - & + & + \\
V2 & + & + & + & + & + & + & + & + & + & - & - & - & - & + \\
V3 & + & + & + & + & + & + & + & + & + & - & - & - & + & + \\
V4 & + & - & + & - & + & - & + & + & + & - & - & - & + & + \\
V5 & + & - & + & - & + & - & + & - & + & - & - & - & + & + \\
V6 & + & + & + & + & + & - & + & + & + & - & - & - & + & + \\
V7 & + & + & + & + & + & + & + & + & + & - & + & - & - & - \\
V8 & + & + & + & + & + & + & + & + & + & - & + & + & + & + \\
V9 & + & + & + & + & + & + & + & + & + & + & - & - & - & + \\
V10 & + & + & + & + & + & + & + & + & + & - & - & - & - & + \\
V11 & + & + & + & + & + & + & + & + & + & - & - & - & + & + \\
V12 & + & + & + & + & + & + & + & + & + & + & - & + & + & + \\
V13 & + & - & + & + & + & - & + & + & + & - & - & - & + & + \\
V14 & + & + & + & + & + & + & + & + & + & - & - & - & + & + \\
V15 & + & + & + & + & + & + & + & + & + & - & + & - & - & + \\
V16 & + & + & + & + & + & - & + & + & + & - & - & - & + & + \\
V17 & + & - & + & - & + & - & + & + & + & - & - & - & + & + \\
V18 & + & + & + & + & + & + & + & + & + & + & + & + & - & + \\
V19 & + & + & + & + & + & + & + & + & + & + & + & - & - & + \\
V20 & + & + & + & + & + & + & + & + & + & - & - & - & - & + \\
V21 & + & + & + & + & + & + & + & + & + & - & - & - & - & + \\
V24 & + & - & + & - & + & - & + & + & + & - & - & - & + & + \\
V25 & + & + & + & + & + & - & + & + & + & - & - & - & - & - \\
V26 & + & - & + & - & + & - & + & + & + & - & - & - & - & - \\
V27 & + & - & + & + & + & - & + & + & + & - & - & - & + & + \\
V28 & + & + & + & + & + & + & + & + & + & + & + & + & + & + \\
V29 & + & + & + & + & + & + & + & + & + & - & - & - & - & + \\
V30 & + & + & + & + & + & + & + & + & + & - & + & - & - & + \\
V31 & + & + & + & + & + & + & + & + & + & + & + & + & - & + \\
V32 & + & + & + & + & + & + & + & + & + & + & + & + & - & + \\
V33 & + & + & + & + & + & + & + & + & + & - & - & - & + & + \\
V34 & + & + & + & + & + & - & + & + & + & - & - & - & + & + \\
V35 & + & + & + & + & + & + & + & + & + & - & - & - & + & + \\
V36 & + & - & + & + & + & - & + & - & + & - & - & - & + & - \\
V37 & + & - & + & - & + & - & + & + & + & - & - & - & - & - \\
V38 & + & + & + & + & + & - & + & + & + & - & - & - & + & + \\
V39 & + & + & + & + & + & + & + & + & + & - & - & - & - & + \\
V40 & + & + & + & + & + & + & + & + & + & - & - & - & + & + \\
V41 & + & + & + & + & + & + & + & + & + & - & - & - & - & + \\
V43 & + & + & + & + & + & + & + & + & + & - & - & - & + & + \\
V44 & + & + & + & + & + & + & + & + & + & - & - & - & + & + \\
V45 & + & + & + & + & + & - & + & + & + & - & - & - & + & + \\
V47 & + & + & + & + & + & - & + & + & + & - & - & - & + & + \\
V52 & + & + & + & + & + & - & + & + & + & - & - & - & + & + \\
V54 & + & - & + & - & - & - & - & - & + & - & - & - & + & + \\
V55 & + & + & + & + & + & + & + & + & + & - & + & - & - & + \\
V56 & + & - & + & - & + & - & + & + & + & - & - & - & + & + \\
V57 & + & - & + & - & + & - & + & + & + & - & - & - & + & + \\
V58 & + & + & + & + & + & + & + & + & + & - & - & - & - & - \\
V59 & + & + & + & + & + & + & + & + & + & + & - & + & + & + \\
V60 & + & - & + & - & + & - & + & + & + & - & - & - & - & - \\
V61 & + & + & + & + & + & + & + & + & + & - & - & - & - & + \\
V62 & + & + & + & + & + & + & + & + & + & - & + & - & - & + \\
V63 & + & + & + & + & + & + & + & + & + & - & - & - & - & + \\
V64 & + & + & + & + & + & + & + & + & + & - & - & - & - & + \\
V65 & + & + & + & + & + & + & + & - & + & - & - & - & + & + \\
V66 & + & + & + & + & + & + & + & + & + & - & - & - & - & + \\
V67 & + & + & + & + & + & + & - & - & + & - & - & - & - & - \\
V68 & + & + & + & + & + & + & - & + & + & - & - & - & - & - \\
V69 & + & + & + & + & + & + & - & + & + & - & - & - & - & - \\
\end{tabular}
\end{minipage}
\end{table*}

\begin{table*}
\begin{minipage}{\textwidth}
\contcaption{}
\begin{center}
\begin{tabular}{ccccccccccccccc}
\hline
 Star & B17 & S27 & O41 & C69 & B24 & K71 & A48 & L24 & P40 & C87 & S91 & C92 & R96 & K00 \\
\hline
V70 & + & + & + & + & + & + & - & + & + & - & - & - & - & - \\
V71 & + & + & + & + & + & + & - & + & + & - & - & - & - & - \\
V72 & + & + & + & + & + & + & - & + & + & - & - & - & - & - \\
V73 & + & + & + & + & + & + & + & + & + & - & - & - & - & + \\
V74 & + & + & + & + & + & + & + & + & + & - & - & - & - & + \\
V75 & + & + & + & + & + & + & + & + & + & - & - & - & - & + \\
V76 & + & + & + & + & + & + & + & + & + & - & - & - & - & + \\
V77 & + & + & + & + & + & + & + & + & + & - & - & - & - & + \\
V78 & + & + & + & + & + & + & + & + & + & - & - & - & + & + \\
V79 & + & + & + & + & + & + & + & + & + & + & - & + & + & + \\
V80 & + & + & + & + & + & - & + & + & + & - & - & - & + & + \\
V81 & + & + & + & + & + & + & + & + & + & - & - & - & + & + \\
V82 & + & - & + & - & - & - & - & + & + & - & - & - & + & + \\
V83 & + & + & + & + & - & - & - & + & + & - & - & - & + & - \\
V85 & + & - & + & - & - & - & - & - & + & - & - & - & - & - \\
V87 & + & + & + & + & + & + & + & + & + & - & - & - & + & + \\
V88 & + & - & + & - & - & - & - & + & + & - & - & - & + & + \\
V89 & + & - & + & - & - & - & - & + & + & - & - & - & + & + \\
V90 & + & - & + & - & - & - & - & - & + & - & - & - & - & - \\
V91 & + & - & + & - & - & - & - & - & + & - & - & - & - & - \\
V92 & + & + & + & + & + & - & - & + & + & - & - & - & - & - \\
V95 & - & - & + & - & - & - & - & - & + & - & - & - & - & - \\
V96 & - & - & + & - & - & - & - & - & + & - & - & - & + & - \\
V97 & - & - & + & - & - & - & - & - & + & - & - & - & + & - \\
V98 & - & - & + & + & - & - & - & - & + & - & - & - & - & - \\
V99 & - & - & + & - & - & - & - & - & + & - & - & - & + & - \\
\hline
\end{tabular}
\end{center}
References: B17: \citet{ba17}; S27: \citet{sh27}, \citet{cs69}, \citet{co71b}; O41: \citet{oo41}; C69: \citet{cs69}; B24: Budapest 60-cm telescope, Table~\ref{mag}; K71: \citet{kk71}; A48: Asiago 1.2-m telescope, Table\ref{mag}; L24: Las Campas 61-cm telescope, Table\ref{mag}; P40: Piszk\'estet\H{o} 1-m telescope, Table\ref{mag}; C87 \citet{cg87}; S91: \citet{st91}; C92: \citet{cm92}; R96: \citet{re96}; K00: \citet{ka00}
\end{minipage}
\end{table*}

\begin{table*}
\caption{Phase shifts and instantaneous periods for V1. The complete table, including all the variables, is given in the electronic version of the article as Supporting Information (Table S3).}
 \label{stars}
  \begin{tabular}{lcrrcl}
  \hline
V1& \multicolumn{2}{c}{  $P_{\mathrm{a}}$=0.5217868 d} &\multicolumn{2}{c}{LC JD2444750--2446209$^{a}$}&{rms = 0.136$^{b}$}\\
 \hline
time interval [JD$-2400000$]&     $\overline {JD}$  & N &     $O-C$ [d]&      error (1$\sigma$)&\\
\hline
       11145.8 -- 14437.6& 13570.6& 80     & -0.00084 &       0.00275&\\
       14795.6 -- 21435.7& 20387.3& 136    & -0.00429 &       0.00163&\\
       27540.8 -- 31259.6& 28533.9& 128    & -0.00134 &       0.00290&\\
       31969.7 -- 34929.6& 34124.7& 204    & -0.00449 &       0.00114&\\
       35251.5 -- 37116.6& 36338.0& 141    & -0.00128 &       0.00225&\\
       37436.5 -- 39979.4& 38900.4& 193    & -0.00186 &       0.00282&\\
       40985.6 -- 43286.6& 42383.3& 267    & 0.00060  &      0.00086&\\
       43631.6 -- 45472.5& 44583.2&287     &-0.00004  &      0.00071&\\
       45850.6 -- 48013.9& 46684.3&140     &-0.00244  &      0.00099&\\
       48392.8 -- 50674.5& 49140.6&778     &-0.00639  &      0.00051&\\
\hline
time interval [JD$-2400000$]&     $\overline{JD}$  & N&     $P$ [d]      & error (1$\sigma$)&    $(P-P_{\mathrm{a}})*10^{5}$d\\
\hline
      11145.8 -- 14437.6 &13570.6 & 80&0.5217821&      .0000023&     -0.47\\
      14795.6 -- 21435.7 &20387.3 &136&0.5217864&      .0000003&     -0.04\\
      27540.8 -- 31259.6 &28533.9 &128&0.5217873&      .0000006&     \,\,0.05\\
      31969.7 -- 34929.6 &34124.7 &204&0.5217877&      .0000007&     \,\,0.09\\
      35251.5 -- 37116.6 &36338.0 &141&0.5217860&      .0000023&    -0.08\\
      37436.5 -- 39979.4 &38900.4 &193&0.5217867&      .0000017&    -0.02\\
      40985.6 -- 43286.6 &42383.3 &267&0.5217873&      .0000006&    \,\,0.05\\
      43631.6 -- 45472.5 &44583.2 &287&0.5217858&      .0000008&    -0.10\\
      45850.6 -- 48013.9 &46684.3 &140&0.5217863&      .0000012&    -0.05\\
      48392.8 -- 50674.5 &49140.6 &778&0.5217858&      .0000002&    -0.10\\

\hline
\multicolumn{6}{l}{\footnotesize{$^{a}$ data used to define the normal light curve; LC and PI refer to the Las Campanas and Piszk\'estet\H{o} observations, respectively.}}\\
\multicolumn{6}{l}{\footnotesize{$^{b}$ residual scatter of the time-transformed data with the exclusion of CCD $V$ data.}}

\end{tabular}
\end{table*}

We consider only those RR Lyrae stars of M5 that have a long enough historical record to follow their period changes i.e. variables with numbering below 100. V53, V86, V93 and V94 are omitted because their photometry is seriously affected by the contamination of close companions. In all, we investigate the period changes of 65 RRab and 21 RRc stars of M5.

We used the original photometric data (reviewed in Sections 2.1 and 2.2) to derive the $O-C$s. Table~\ref{data} summarizes the available photometry for each RR Lyrae star of M5, the sources are denoted by `+'.

The procedure applied to the determination of the $O-C$ diagrams and to explore the period changes is basically the same as used by \cite{ocen} in the study of $\omega$ Centauri's variables.

As the first step, the zero-points of the  magnitudes of the different observations of each RR Lyrae star were homogenized. The mean light curves used for the magnitude transformations were the Las Campanas data from 1981--1991 and  for the 14 stars (V5, V36, V54, V65, V67, V85, V87, V90, V91, V95, V96, V97, V98, V99) that were not or only partly measured on the Las Campanas plates, they were constructed from the 1976--1993  Piszk\'estet\H{o} data. The light curves of the RRab and RRc stars have been decomposed by fifth and third order Fourier series, respectively, to fix the mean light curves. For the sake of uniformity, we used these relatively low-order fits for the RRab stars, because higher-order solutions gave wavy curves for some of the stars. We checked, however, to ensure that the zero-points and phase shifts would remain unchanged if higher order fits had been used for the variables with good-quality light curves.

Both vertical and horizontal shifts were determined for all the different photometric data sets of the variables in order to match their normal curves best. The first part of the Las Campanas observations, from 1972 to 1980 were treated as a separate data set because of the zero-point inconsistencies mentioned in Sect. 2.2. Then, the derived vertical (zero-point) shifts have been applied to each data set of each star. In this way, consistent data sets have been obtained for the variables. The most outlying magnitudes were omitted from some of the data sets.

For homogeneity purposes the $B$ band CCD observations were used if they existed. The CCD observations of \citet{cm92}, \citet{re96} and \citet{ka00} were in $V$ (and $I$) bands. It is well-known that the times of maximum systematically deviate in blue and yellow light, but \cite{ocen} have shown if the phase shifts of the whole light curve are considered (as in our case), no significant difference according to the colours occurs. Therefore, CCD $V$ data were used and treated in the same manner as photographic and CCD $B$ observations. A similar argument holds for the inhomogeneous photographic and CCD $B$ data. Although the amplitudes of the light curves and the phases of the maximum times might be somewhat different in the different observations, this had a negligible effect on the derived phase-shift values that were used for the $O-C$ analysis.

As the second step, the $O-C$ values were determined. They have been calculated as the horizontal shifts of different parts of the combined, magnitude zero-point homogenized data sets to the normal curves. In order to eliminate smearing of the light curves caused by period changes, the $O-C$s should be constructed only from 3--5 years of the observations.

The normal curves that define the zero value of the $O-C$ data have been constructed using 100--180 data points. Each was checked visually for complete phase coverage. The 1981--1985 Las Campanas observations were used to construct the normal curve for most of the variables. For stars with rapid period variations (V14, V18, V24, V25, V52, V66, V70, V81 and V92) an even shorter time interval (1981--1983) has been employed, while data from 1982--1983 defined the normal curve of V76. For the 14 stars not or only partly (V87) measured on the Las Campanas plates, the normal curves were constructed from the 1976--1980 Piszk\'estet\H{o} data.

$O-C$ values were calculated, as a rule, for ten groups (subsets) of the whole, zero-point corrected data sets of each RR Lyrae star. In the case of stars for which measurements were not available for all the extended data sources a smaller number of groups were composed. Also, if the phase shift could only be poorly determined, or the scatter of the light curve was too large, groups were merged, in order to stabilize the solution. Conversely, if the period changed too fast and/or irregularly, more than ten groups were formed and analysed. The $O-C$ values deduced from the observations of \cite{br96}  outlie significantly in the $O-C$ plots; therefore, they were disregarded in the phase diagrams.

As starting periods, we accepted the values given in the literature (in some cases frequency analysis was needed) and constructed preliminary $O-C$ diagrams. Then, the average period values ($P_{\mathrm{a}}$) over the century were determined and used in order to obtain a symmetrical form of the diagrams. If it was necessary, several trials were done to find the best period, which gave the most reasonable $O-C$ plot. This was especially the case, if significant abrupt and irregular period variations occurred. 

If the $O-C$ is fitted by a polynomial 
\begin{equation}
O-C=\sum_{i=1}^{k} c_{i} t^{i-1}
\end{equation}
then it can be readily seen that the period variation can be eliminated, if the times of the observations are transformed according to the equation
\begin{equation}
t'=t-\sum_{i=1}^{k} c_{i} t^{i-1}.
\end{equation}
The time-transformed data can be coherently phased  with the period, $P_{\mathrm{a}}$. 

For some of the variables with strong period variations, the zero-point corrections of the different data sets had slightly incorrect values because of the smearing of the folded light curves. In such cases, we used the time-transformed light curves to derive refined values for the corrected magnitudes.

In summary, the following steps were applied:

\begin{itemize}
\item{$m_0,t_0 \rightarrow m_1, t_0$: magnitude shift of the different observations,}
\item{$m_1,t_0 \rightarrow m_1, t_1$: time transformation of the data using equation 2,}
\item{$m_1,t_1 \rightarrow m_2, t_0$: refining the magnitude transformation on the time-transformed data,}
\item{$m_2,t_0 \rightarrow m_2, t_2$: time transformation of the refined magnitude data sets,}
\end {itemize}
where $m_0$, $t_0$, $m_1$, $t_1$ and  $m_2$, $t_2$ refer to the magnitudes and times of the original observations; the zero-point corrected, time-transformed data; and the refined, zero-point corrected, time-transformed data.

All the obervations used in the analysis and the magnitude-zero-point and time-transformed data are given in electronic form as Supporting Information (Table S4) and can be downloaded from the url: {http://www.konkoly.hu/24/publications/M5}. Table~\ref{all} gives a sample of the
data.

\begin{table}
\caption{The collected photometric observations of M5 variables. The complete table is given in the electronic version of the article as Supporting Information (Table S4). }
 \label{all}
  \begin{tabular}{lccccc}
  \hline
HJD$-2400000$ & mag & source$^{a}$&  mag$_{\mathrm {tr}}^{b}$ &HJD$_{\mathrm {tr}}^{c}$& star\\
 \hline
11145.794 & 14.46 & B17  & 14.72  & 11145.794 &  V1 \\
...&...&...&...&...&...\\
21338.885 & 14.35 &  S27  & 14.48 &  21338.885 &  V1 \\
...&...&...&...&...&...\\
11178.785  & 14.94 &  B17  & 15.38  & 11178.708  & V2 \\
...&...&...&...&...&...\\
\hline
\multicolumn{6}{l}{\footnotesize{$^{a}$ the same as in Table~\ref{data}}}\\
\multicolumn{6}{l}{\footnotesize{$^{b}$ zero-point transformed, homogenized magnitudes}}\\
\multicolumn{6}{l}{\footnotesize{$^{c}$ transformed HJD that eliminates phase/period variations}}\\
\hline
\end{tabular}
\end{table}

\section{$O-C$ diagrams, period changes and folded light curves}

\begin{figure*}
\includegraphics[angle=-90,width=17.7cm]{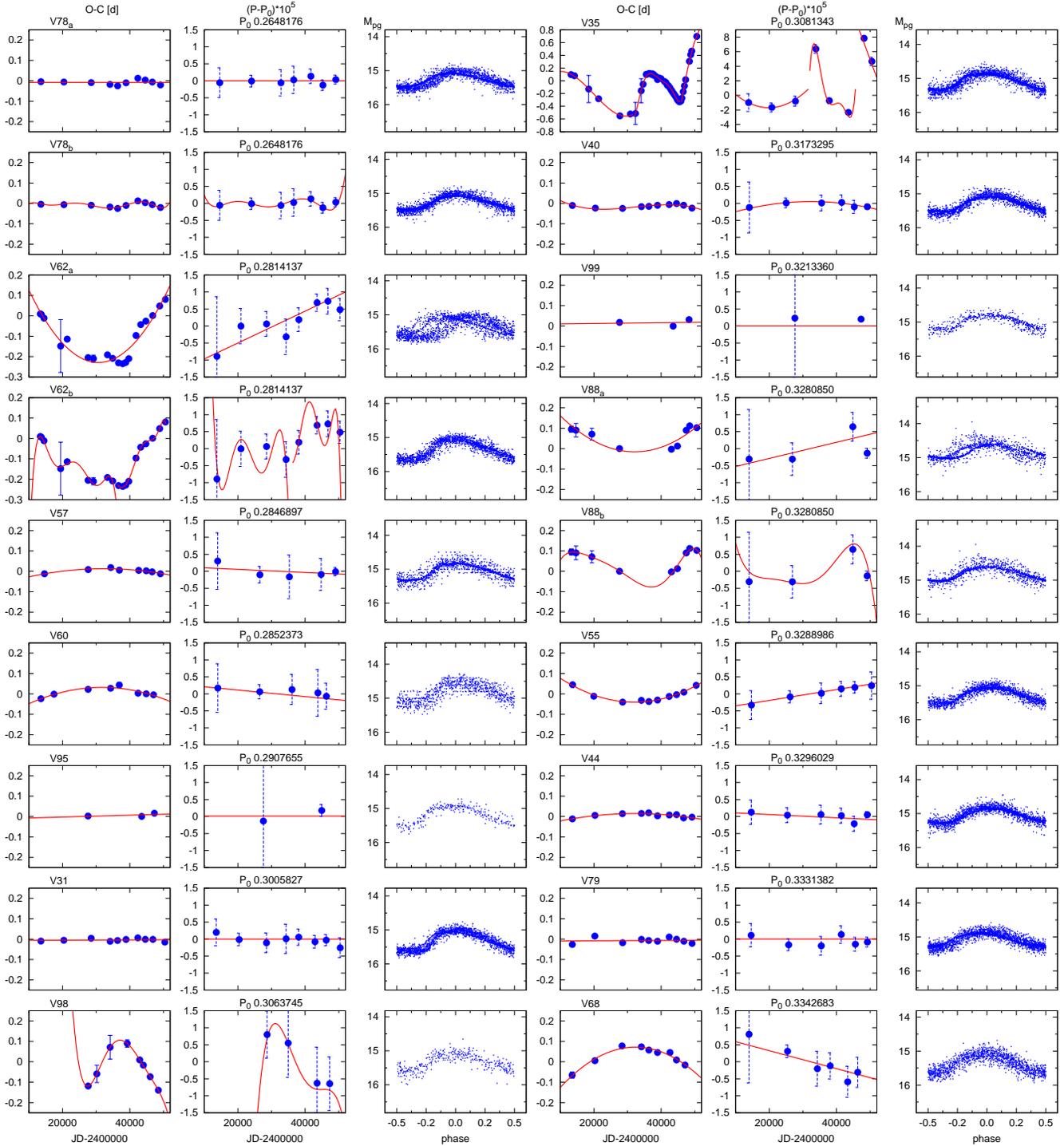}
\caption{For each RR Lyrae star, in increasing order of the period, three panels describe the behaviour of the period changes. The left-hand panels show the $O-C$ diagrams and their polynomial fits. The $O-C$ values are given in days, and are plotted versus Julian Date. In the middle panels the departures from the average values of the period  [$(P-P_{\mathrm{a}})\times 10^5$ d] are plotted from direct period determination for the data subsets versus Julian Date. $2\sigma$ formal errors of the $O-C$ and direct period values are indicated. For comparison, the derivatives of the $O-C$ polynomial fits are also drawn in. The right-hand panels show the folded light curves of the time-transformed data calculated from the $O-C$ polynomial fits over the hundred years of observations. The folded light curves give an impression about the quality of the $O-C$ fits.}
\label{oc}
\end{figure*}
\begin{figure*}
\includegraphics[angle=-90,width=17.7cm]{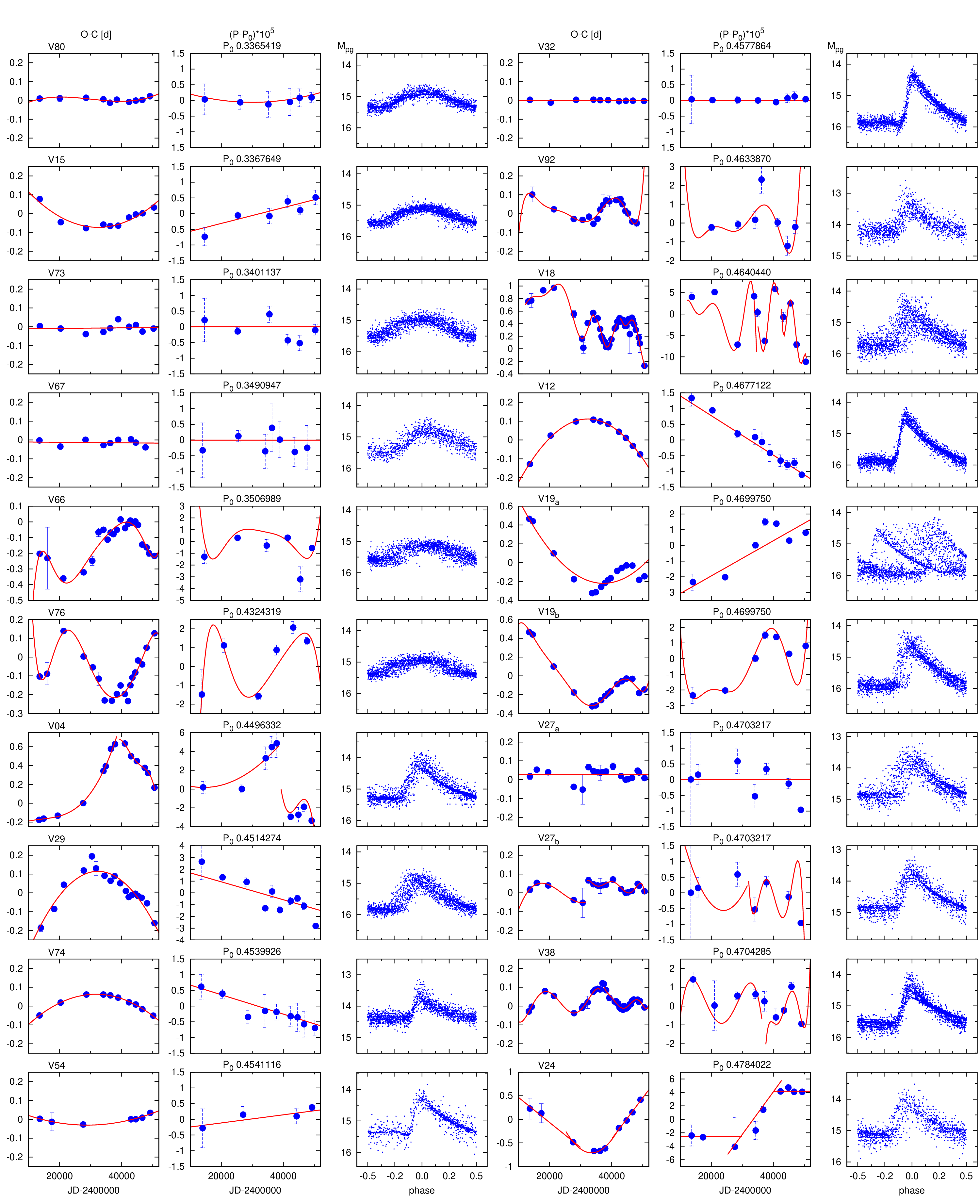}
\contcaption{}
\end{figure*}
\begin{figure*}
\includegraphics[angle=-90,width=17.7cm]{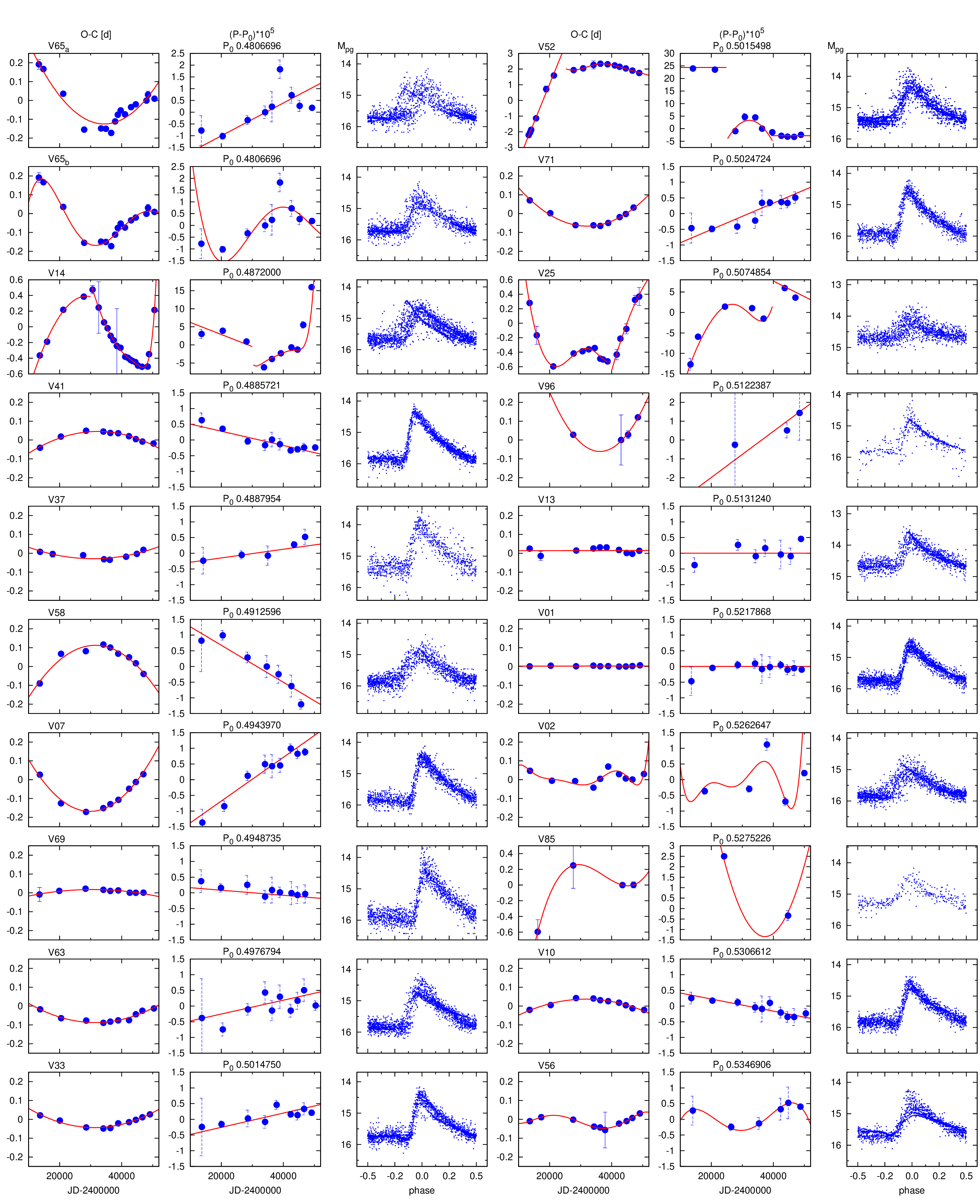}
\contcaption{}
\end{figure*}
\begin{figure*}
\includegraphics[angle=-90,width=17.7cm]{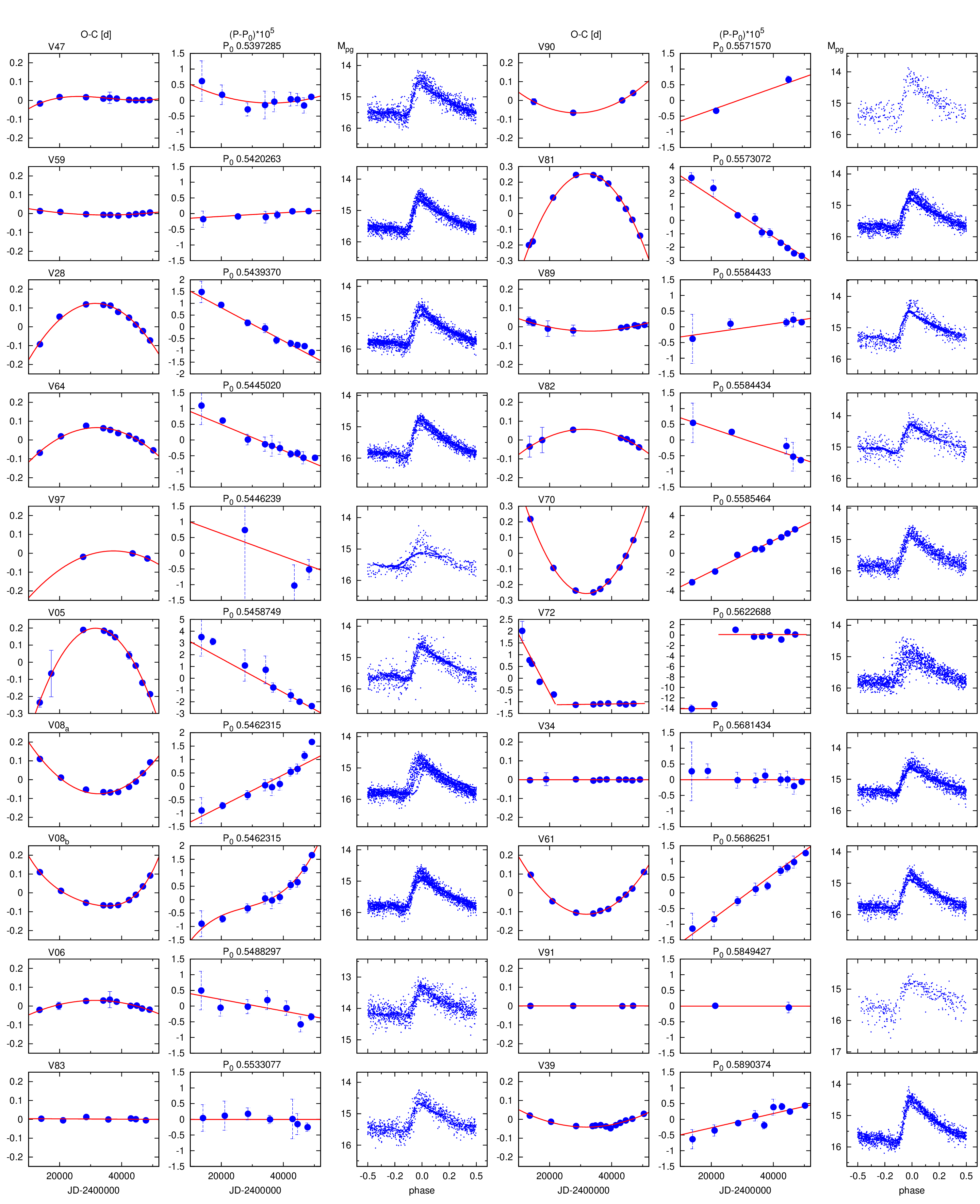}
\contcaption{}
\end{figure*}
\begin{figure*}
\includegraphics[angle=-90,width=17.7cm]{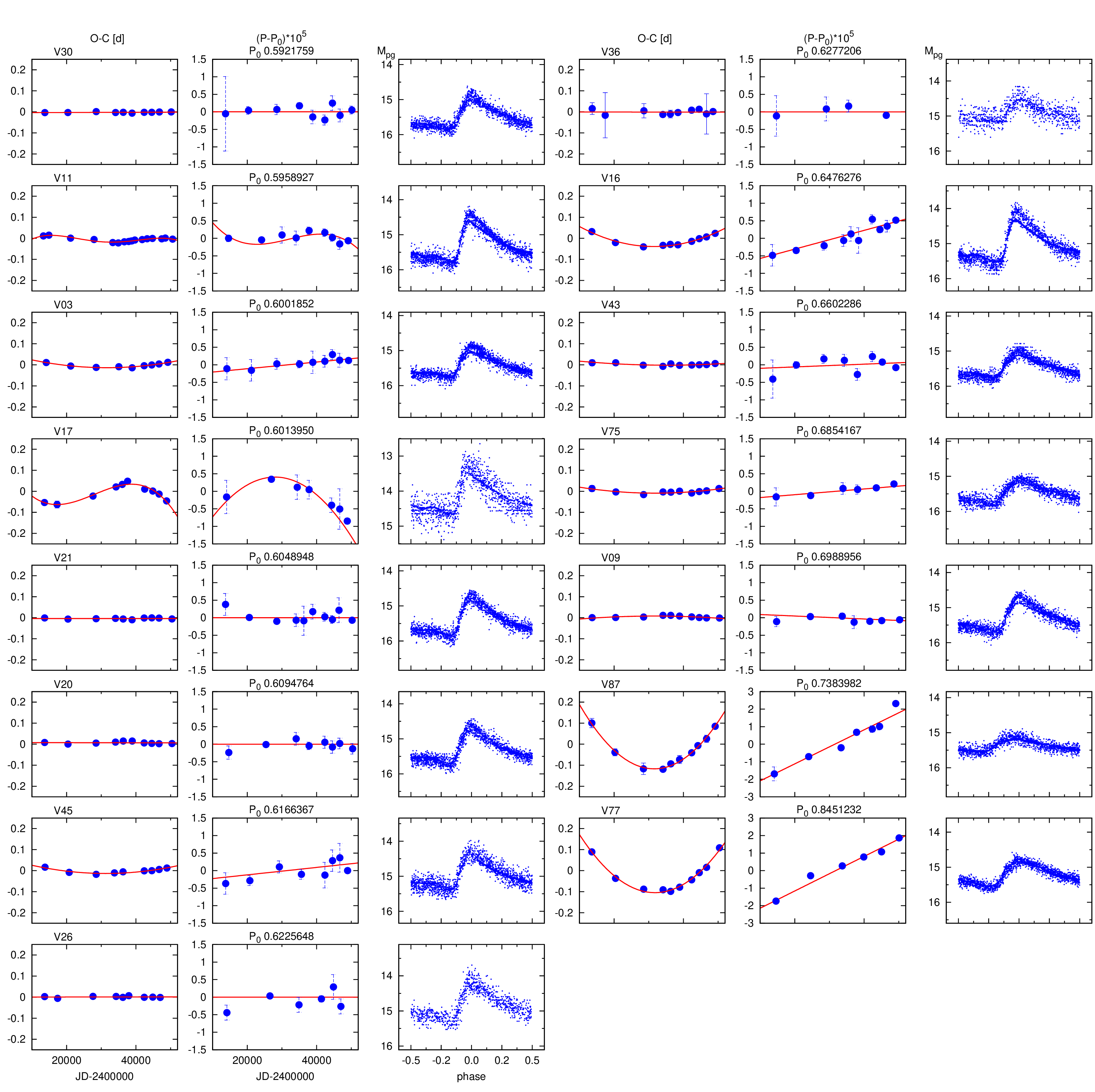}
\contcaption{}
\end{figure*}

The results for the 21 RRc and 65 RRab stars are documented in Fig.~\ref{oc}. Three panels are shown for each variable. The $O-C$ values and their errors, estimated as twice the standard deviation of the corresponding part of the light curve from the light-curve template used to determine the $O-C$ value, are plotted versus time in the left-side panels. For the sake of uniformity and simplicity, the $O-C$ diagrams have been approximated by polynomials of different order. Whenever it was possible, a linear or quadratic fit was applied. Occasionaly, when the period changed very irregularly, the $O-C$ diagram was composed from two or three polynomials fitted to the different parts of the $O-C$ diagram. In these cases, the $O-C$ solution might not be unambiguous because of cycle-count uncertainties.

In the middle panels, the deviations from the average values of the periods $(P-P_{\mathrm{a}})\times10^5$ are plotted from direct period determination for the data subsets. 2$\sigma$ formal errors of the direct period determinations are also shown. If the period could be determined only with large uncertainty, data subsets were drawn together. 

The $O-C$ can be given as the function of time as
\begin{equation}
O-C={1\over P_{a} }\int P(t) {\mathrm d}t - t.
\end{equation}
Consequently, the temporal period, $P(t)$, can be approximated using the $c_i$ coefficients of Eq. 1.:  
\begin{equation}
P(t)=P_{a}{{{\mathrm d}(O-C)}\over{{\mathrm d}t}}+P_{a}=P_{a}\sum_{i=2}^{k} (i-1) c_{i} t^{i-2} +P_{a}.
\end{equation}

For comparison, and to confirm the $O-C$ solution, $P(t)$ calculated from the derivatives of the polynomial $O-C$ fits are also drawn in the middle panels of Fig~\ref{oc}.

The right-hand panels show the folded light curves of the time-transformed data (Eq. 2.) calculated from the $O-C$ polynomial fits over the hundred years of observations.

Two $O-C$ solutions are shown for some of the stars (V8, V19, V27, V38, V62, V65, V78 and V88). The first fits the $O-C$ data with linear or parabolic approximation, while the second shows a higher order or multiple fit. These examples illustrate the strengths of the higher order polynomial or irregular period variations.
We also note here that, in some cases (V11, V17, V56 RRab and V40, V80 RRc stars), a sine-like approximation would also be appropriate, but for the sake of conformity and uniformity, as mentioned, a polynomial fit was always applied.

Although a visual inspection of the folded light curve was already convincing of the validity of both the $O-C$ values and their polynomial fits in most cases, we also checked this  quantitatively. The residuals of the Fourier fits of the folded light curves (omitting CCD $V$ data) of the time-transformed data have been determined and compared to the mean rms of the observations. For variables with period variation that was not too complex, the rms of the time-transformed data were 0.10--0.15 mag for well resolved stars, which is in the same range as the typical errors of the photographic observations. The rms of the photographic observations of these stars ranged from 0.05 to 0.20-mag (usually the smallest residuals were obtained from iris photometry) with an average value of about 0.10--0.15 mag.

The numerical results for the $O-C$ and period data of the studied 86 variables are available electronically. Table~\ref{stars} shows an example for V1. The first line of the table identifies the variable, gives the period that was used to calculate the $O-C$, the data sets used for constructing the normal curve and the rms of the time-transformed light curve. In calculating this rms, only photographic and $B$ band observations were used. The first part of Table~\ref{stars} lists the $O-C$ data. The time intervals, their mean Julian Date, the number of data points, the $O-C$ value and its 1$\sigma$ error are given. The second part of Table~\ref{stars} gives similar data for direct period determinations. The last column of this part of the table lists the actual period-deviation values, $P-P_{\mathrm{a}}$.

We emphasize here that the $P_{\mathrm{a}}$ periods are not the currently best periods of the variables; these periods are the best mean periods over the last century. The periods determined for the latest part of the observations match the recent data.

Amplitude and phase modulations of Blazhko stars might introduce some uncertainty into the derived zero-point shifts and $O-C$ values in the case of variables showing strong light-curve modulations. However, most of the Blazhko stars identified in \cite{p2} show strongly irregular period variations, which cannot be used to derive any period-change rate. Therefore, these uncertainties have no effect on any of the conclusions of this paper. The period-change rates of the two Blazhko stars with strong phase modulations (V58 and V63), that have dominantly continuous period changes are well defined as can be seen in Fig.~\ref{oc}. We have to admit, however, that the origin of the scatter of the $O-C$ and period variations of these stars may either originate from the bias of their phase modulation or may reflect real small fluctuations in the periods.

\section{Results and Discussion}

\begin{table*}
\caption{Summary of the light-curve parameters and period-change properties of M5 RR Lyrae stars. }
\begin{minipage}{\textwidth}
\label{tabla}
\begin{tabular}{lccccclcccrrrr}
Star &type& $P_{\mathrm{a}}^a$ & $V_i^b$ & $V_m^b$ & $B_m^b$&$A_V^c$ & \multicolumn{3}{c}{Remarks}& $\beta=\dot{P}$ & $\beta=\dot{P}$& $\alpha=P^{-1}\dot{P}$ & $P_{\mathrm{max}}-P_{\mathrm{min}}^g$ \\
\hline
& & d &\multicolumn{4}{c}{mag} & $^d$& $^e$&$^f$  & $10^{-10}$ & $\mathrm{d Myr^{-1}}$ & $\mathrm{10^{-10}d^{-1}}$ &  $10^{-5}$d\\
\hline
V1& RRab&0.5217868&15.103&15.155&15.443&1.07--1.15&0 &0&Bl& 0.00&0.00   &0.00  & 0.56\\
V2& RRab&0.5262647&15.093&15.148&15.559&1.04--1.11&--&2&Bl& --  & --    & --   & 1.82\\
V3& RRab&0.6001852&15.057&15.078&15.451&0.74     &1  &0&  & 0.80&  0.029&  1.32& 0.45\\
V4& RRab&0.4496332&15.047&15.093&15.343&0.95--1.25&--&2&Bl& --  &   --  &   -- & 8.22\\
V5& RRab&0.5458749&15.107&15.165&15.390&1.02--1.18&1 &0&Bl:&-13.11&-0.479&-24.02&5.88\\
V6& RRab&0.5488297&15.033&15.076&   -- &1.00     &1 &0& d& -1.68& -0.061& -3.06& 1.08\\
V7& RRab&0.4943970&15.035:&15.103:&15.491:&1.30  &1 &0&  &  7.04&  0.257& 14.23& 2.36\\
V8& RRab&0.5462315&15.085&15.126&15.497&0.82--1.08&--&1&Bl& --  &  --   &  --  & 2.55\\
V9& RRab&0.6988956&14.903&14.931&15.324&0.80     &1 &0&  & -0.30& -0.011& -0.43& 0.17\\
V10&RRab&0.5306612&15.085&15.133&15.516&1.10     &1 &0&  & -1.81& -0.066& -3.41& 0.61\\
V11&RRab&0.5958927&14.933&14.988&15.345&1.13     &--&1&  &  --  &  --   & --   & 0.38\\
V12&RRab&0.4677122&15.098&15.172&15.502&1.28     &1 &0&  & -6.72& -0.245&-14.37& 2.44\\
V13&RRab&0.5131240&14.955&15.006&   -- &1.15     &0 &1& d& 0.00 &0.00   &0.00  & 0.83\\
V14&RRab&0.4872000&15.080&15.131&15.424&0.50--1.35&--&2&Bl&  -- &   --  &   -- &22.12\\
V15&RRc &0.3367649&15.054&15.064&15.355&0.41     &1 &0&  &  3.74&  0.137& 11.11& 1.25\\
V16&RRab&0.6476276&14.809&14.869&15.267&1.19     &1 &0&  &  2.06&  0.075&  3.19& 1.03\\
V17&RRab&0.6013950&14.942&14.997&   -- &1.12     &2 &1& d&  --  &   --  &  --  & 1.19\\
V18&RRab&0.4640440&15.075&15.151&15.420&0.80--1.35&--&2&Bl&  -- &   --  &   -- &17.00\\
V19&RRab&0.4699750&15.088&15.173&15.495&1.25--1.35&--&2&Bl&  -- &   --  &   -- & 3.84\\
V20&RRab&0.6094764&15.024&15.058&15.429&0.91     &0 &0&  & 0.00 & 0.00  & 0.00 & 0.39\\
V21&RRab&0.6048948&15.004&15.045&15.441&0.98     &0 &0&  & 0.00 & 0.00  & 0.00 & 0.48\\
V24&RRab&0.4784022&15.068&15.100&15.363&0.85--0.95&--&2&Bl&  -- &   --  &   -- & 8.79\\
V25&RRab&0.5074854&   -- &   -- &   -- &  --     &--&2& d&   -- &   --  &   -- &18.63\\
V26&RRab&0.6225648&   -- &   -- &   -- &  --     &0 &0& d& 0.00 & 0.00  & 0.00 & 0.73\\
V27&RRab&0.4703217&14.995&15.053&15.356&1.12--1.50&--&2&Bl&  -- &   --  &   -- & 1.55\\
V28&RRab&0.5439370&15.087&15.123&15.511&0.94     &1 &0&  & -6.45& -0.236&-11.86& 2.57\\
V29&RRab&0.4514274&15.127&15.164&15.506&0.88--0.88&1 &1&Bl:&-8.01&-0.292&-17.73& 5.46\\
V30&RRab&0.5921759&15.067&15.093&15.516&0.82--0.90&0 &0&Bl&0.00 & 0.00  & 0.00 & 0.48\\
V31&RRc &0.3005827&15.053&15.069&15.298&0.50     &0 &0&  & 0.00 & 0.00  & 0.00 & 0.45\\
V32&RRab&0.4577864&15.075&15.146&15.480&1.31     &0 &0&  & 0.00 & 0.00  & 0.00 & 0.18\\
V33&RRab&0.5014750&15.073&15.129&15.551&1.14     &1 &0&  &  2.26&  0.083&  4.51& 0.71\\
V34&RRab&0.5681434&15.041&15.067&15.426&0.82     &0 &0&  & 0.00 & 0.00  & 0.00 & 0.48\\
V35&RRc &0.3081343&14.956&14.970&15.225&0.48     &--&2&  &   -- &  --   &  --  &10.17\\
V36&RRab&0.6277206&15.059&15.080&15.530&0.65     &0 &0& d& 0.00 & 0.00  & 0.00 & 0.28\\
V37&RRab&0.4887954&   -- &   -- &   -- &  --     &1 &0& d&  1.41&  0.052&  2.89& 0.76\\
V38&RRab&0.4704285&15.090&15.138&15.484&0.85--1.03&--&2&Bl&  -- &   --  &   -- & 2.36\\
V39&RRab&0.5890374&14.944&14.999&15.386&1.17     &1 &0&  &  1.88&  0.069&  3.19& 1.07\\
V40&RRc &0.3173295&15.040&15.051&15.304&0.43     &2 &1&  &   -- &   --  &   -- & 0.15\\
V41&RRab&0.4885721&15.085&15.138&15.466&1.06     &1 &0&  & -2.32& -0.085& -4.76& 0.96\\
V43&RRab&0.6602286&15.031&15.047&15.478&0.62     &1 &0&  &  0.30&  0.011&  0.45& 0.65\\
V44&RRc &0.3296029&15.023&15.034&15.253&0.43     &1 &0&  & -0.73& -0.027& -2.22& 0.34\\
V45&RRab&0.6166367&14.979&15.015&15.480&1.00     &1 &0&  &  0.86&  0.031&  1.40& 0.74\\
V47&RRab&0.5397285&15.101&15.147&15.489&1.04     &--&1&  &   -- &   --  &   -- & 0.90\\
V52&RRab&0.5015498&14.976&15.029&15.337&1.00--1.20&--&2&Bl&  -- &   --  &   -- &27.26\\
V54&RRab&0.4541116&15.012&15.078&   -- &1.28     &1 &0& d&  1.43&  0.052&  3.14& 0.66\\
V55&RRc &0.3288986&15.075&15.085&15.358&0.41     &1 &0&  &  2.42&  0.089&  7.37& 0.57\\
V56&RRab&0.5346906&15.116&15.134&15.506&0.60--0.85&--&2&Bl&  --  &  --   &  --  & 0.77\\
V57&RRc &0.2846897&15.047&15.062&15.322&0.53     &1 &0&  & -0.80& -0.029& -2.82& 0.47\\
V58&RRab&0.4912596&   -- &  --  &15.573&0.62--0.92&1&0&Bl& -6.00& -0.219&-12.22& 2.20\\
V59&RRab&0.5420263&14.977&15.015&15.430&0.95     &1 &0&  &  0.54&  0.020&  0.99& 0.26\\
V60&RRc &0.2852373&   -- &  --  &15.234&0.54     &1 &0&  & -1.69& -0.062& -5.92& 0.24\\
V61&RRab&0.5686251&15.062&15.095&15.485&0.91     &1 &0&  &  6.44&  0.235& 11.32& 2.41\\
V62&RRc &0.2814137&15.082&15.099&15.315&0.53     &--&2&  &  --  &  --   & --   & 1.62\\
V63&RRab&0.4976794&15.06:&15.08:&15.454&0.90--1.00&1&1&Bl&  2.22&  0.081&  4.45& 1.25\\
V64&RRab&0.5445020&15.093&15.134&15.563&1.02     &1 &0&  & -3.79& -0.138& -6.96& 1.66\\
V65&RRab&0.4806696&15.085&15.133&15.415&0.75--1.00&--&2&Bl& --  &   --  &   -- & 2.84\\  
V66&RRc &0.3506989&15.019&15.031&15.364&0.44     &--&2&  &  --  &   --  &   -- & 3.53\\
V67&RRc &0.3490947&   -- &  --  &15.137&0.62     &0 &0&  &0.00  & 0.00  & 0.00 & 0.77\\
V68&RRc &0.3342683&   -- &  --  &15.369&0.50     &1 &0&  & -3.96& -0.145&-11.84& 1.40\\
V69&RRab&0.4948735&   -- &  --  &15.515&1.23     &1 &0&  & -0.82& -0.030& -1.65& 0.49\\
\end{tabular}
\end{minipage}
\end{table*}
\begin{table*}
\begin{minipage}{\textwidth}
\contcaption{}
\begin{tabular}{lccccclcccrrrr}
Star &type& $P_{\mathrm{a}}^a$ & $V_i^b$ & $V_m^b$ & $B_m^b$&$A_V^c$ & \multicolumn{3}{c}{Remarks}& $\beta=\dot{P}$ & $\beta=\dot{P}$& $\alpha=P^{-1}\dot{P}$ & $P_{\mathrm{max}}-P_{\mathrm{min}}^g$ \\
\hline
& & d &\multicolumn{4}{c}{mag} & $^d$& $^e$&$^f$  & $10^{-10}$ & $\mathrm{d Myr^{-1}}$ & $\mathrm{10^{-10}d^{-1}}$ &  $10^{-5}$d\\
\hline
V70&RRab&0.5585464&   -- &  --  &15.576&1.08     &1 &0&  & 14.65&  0.535& 26.24& 5.61\\
V71&RRab&0.5024724&   -- &  --  &15.565&1.23     &1 &0&  &  4.19&  0.153&  8.33& 1.00\\
V72&RRab&0.5622688&   -- &  --  &15.525&0.70--0.92&--&2&Bl& --  &   --  &   -- &15.11\\
V73&RRc &0.3401137&14.958&14.974&15.240&0.51     &0 &1&  & 0.00 & 0.00  & 0.00 & 0.92\\
V74&RRab&0.4539926&15.005&15.083&   -- &1.39     &1 &0& d& -3.40& -0.124& -7.49& 1.31\\
V75&RRab&0.6854167&14.989&15.002&15.498&0.55     &1 &0&  &  0.60&  0.022&  0.87& 0.37\\
V76&RRc &0.4324319&14.819&14.827&15.160&0.40     &--&2&  &  --  &   --  &   -- & 3.63\\
V77&RRab&0.8451232&14.760&14.775&15.222&0.60     &1 &0&  &  5.83&  0.213&  6.90& 3.61\\
V78&RRc &0.2648176&15.076&15.086&15.332&0.42     &0 &1&  & 0.00 & 0.00  & 0.00 & 0.26\\
V79&RRc &0.3331382&15.010&15.020&15.251&0.38     &0 &0&  & 0.00 & 0.00  & 0.00 & 0.33\\
V80&RRc &0.3365419&15.042&15.052&15.344&0.41     &2 &1&  &  --  &   --  &   -- & 0.22\\
V81&RRab&0.5573072&15.058&15.094&15.524&0.95     &1 &0&  &-13.57& -0.496&-24.34& 5.80\\
V82&RRab&0.5584434&15.046&15.079&15.508&0.92     &1 &0&  & -3.00& -0.109& -5.36& 1.19\\
V83&RRab&0.5533077&15.092&15.123&15.404&0.84     &0 &0& d&0.00  & 0.00  &0.00  & 0.43\\
V85&RRab&0.5275226&   -- &   -- &   -- &  --     &: &:& d&  --  &  --   &  --  & 2.83\\
V87&RRab&0.7383982&14.915&14.921&15.378&0.37     &1 &0&  &  6.58&  0.240&  8.91& 4.01\\
V88&RRc &0.3280850&15.024&15.034&15.305&0.42     &--&2&  &  --  &  --   &  --  & 0.95\\
V89&RRab&0.5584433&15.085&15.122&15.540&0.96     &1 &0&  &  1.26&  0.046&  2.25& 0.61\\
V90&RRab&0.5571570&   -- &  --  &   -- &  --     &1 &0& d&  3.14&  0.115&  5.64& 0.99\\
V91&RRab&0.5849427&   -- &  --  &   -- &  --     &0 &0& d& 0.00 &  0.00 & 0.00 & 0.06\\
V92&RRab&0.4633870&   -- &  --  &   -- &  --     &--&2& d&  --  &  --   &  --  & 3.52\\
V95&RRc &0.2907655&15.060&15.079&15.300&0.50     &: &:&  &  --  &  --   &  --  & 0.21\\
V96&RRab&0.5122387&15.091&15.136&15.536&0.94     &: &:& d&  --  &  --   &  --  & 1.69\\
V97&RRab&0.5446239&15.042&15.051&15.486&0.40--0.75&:&:&Bl&  --  &  --   &  --  & 1.77\\
V98&RRc &0.3063745&15.120&15.134&15.401&0.52     &--&1&  &  --  &  --   &  --  & 1.44\\
V99&RRc &0.3213360&15.037&15.049&  --  &0.46     &: &:&  &  --  &  --   &  --  & 0.03\\
\hline
\end{tabular}
{\footnotesize
$^{a}$: $P_{\mathrm{a}}$, the period used to construct the $O-C$.\\
$^{b}$: $V_i$, $V_m$ and $B_m$: intensity and magnitude averaged mean magnitudes. \\
$^{c}$ $A_V$: the pulsation amplitude in $V$.\\
 Remarks: $^{d}$: linear (0), quadratic (1), third order (2),
higher order or multiple (--) fit to the $O-C$. Colon denotes stars with scarce $O-C$ data.
$^{e}$: complexity index of the $O-C$ diagram, straight line or parabola (0), low order fit with scatter or third order fit (1), higher order or multiple fit of irregular character (2).
$^{f}$: This column refers only to RRab stars. Blazhko effect (Bl) and deficient photometry for the investigation of light-curve variability (d) are marked.\\
$^{g}$: the full range of the observed period variation.\\
See more details in the text.}

\end{minipage}
\end{table*}

\subsection{Period-change results}
In the previous section the $O-C$ diagrams of 21 RRc and 65 RRab stars of the cluster are presented. The main characteristic parameters of the light curves and period changes of these stars are collected in Table~\ref{tabla}. In the first three columns the catalogue numbers of the variable in question, its type and the period used to construct the $O-C$ diagram are given. 

The next four columns contain some data of the light curves that might have a connection with the stars' evolutionary stage and can be compared with the variables' period-change rates: $V_i$ and $V_m$ are the intensity- and magnitude-averaged mean $V$ magnitudes, respectively, derived from the combined R96 and K00 CCD $V$ observations for most of the stars. The S91 CCD $V$ photometry was used for the star V7 and the \cite{br96} CCD $V$ data for the variables V95 and V98. For Blazhko-stars, light curves of the best-represented phase of the modulation were used to derive the mean $V$ magnitudes. $B_m$ denotes the magnitude-averaged mean $B$ magnitude of the variables determined from the CCD $B$ data or, if they are not available then $B_m$  has been derived from the Las Campanas or the Piszk\'estet\H o photographic observations. No mean B magnitude is given for V6, V13, V17, V25, V26, V37, V54, V74, V85, V90, V91 and V92. There are no CCD $B$ observations of these stars, and their photographic photometry is seriously affected by bright close companions and crowding. The mean photographic magnitudes of these stars are about 0.5--1.5 mag brighter than expected. The mean-magnitude values were calculated as $A_0$, the first member of the Fourier-fit of the light curve of the variables. 

The mean $V$ magnitudes of the variables differed by 0.01--0.03 mag in the R96 and K00 data for the variables observed in both surveys. It is thus estimated that the mean $V$ magnitudes have similar uncertainties. The error of the mean $B$ magnitudes cannot be correctly quantified. Although we have omitted variables with strong light contamination due to crawding, the photographic $B$ magnitudes of some of the variables might still be somewhat biased by their surroundings. The inhomogeneity of the $B$ data (photographic and CCD) introduce further errors. Notwithstanding these problems, the accuracy of the given $B$ magnitudes is estimated to be better than 0.05-mag for most of the stars.

The pulsation amplitude in the $V$ band, $A_V$, has been derived from the CCD $V$ observations that were used to calculate the mean $V$ magnitude. If no $V$ observations were available, the $A_V$ amplitude has been estimated as $A_V \approx A_B/1.3$ \citep{ibvs}, where $A_B$ is the blue amplitude of the light curve derived from the CCD $B$ or photographic $B$ observations. For Blazhko stars, the smallest and largest observed amplitudes, taken from \citet[][Paper II]{p2}, are given.

As the inhomogeneity of the listed mean magnitudes and amplitudes gives rise to enhanced uncertainties, these values have to be taken with caution and are given only for guidance. As no accurate multicolour photometry of the full sample of M5 variables exists, this combined photometric information is, however, also utilized on a statistical ground. 

The `Remarks' supply information on the complexity and quality of the $O-C$ diagram, i.e. on the derived period changes and their reliability for each variable, and on the presence of light-curve variation for RRab stars. The first column of the `Remarks' gives the order of the polynomial fit to the $O-C$ data. The index `0' means straight-line fit, `1' means quadratic, while `2' means cubic approximation. In this column, a `hyphen' indicates that an even higher-order polynomial is fitted (e.g. in case of sine-like $O-C$ variations) or a combined polynomial approximation is applied to satisfactorily describe irregular period variations with sudden changes. In these latter cases, when the $O-C$ cannot be fitted with a single polynomial, cycle-count uncertainties of the $O-C$ solution cannot be excluded. As these $O-C$s are always classified as irregular, i.e. no period-change rate is derived from these data, these ambiguities have no effect on any of our conclusions.

There are five stars (V85, V95, V96, V97 and V98) denoted by colons in this column that refer to their scarcely defined $O-C$ diagrams. These stars are included only for completeness, but the $O-C$ solutions shown in Fig.~\ref{oc} are ill-defined, other solutions are similarly possible. These stars are disregarded in the following.

The second column of the `Remarks' gives information about the complexity of the $O-C$ diagram, i.e. about the irregularity of the period changes. The irregularity indices `1' and `2' refer to moderate and strong irregularity of the $O-C$ diagram while `0' means that the variation in the pulsation period is smooth and regular, i.e. it is either constant or linearly changing. Sometimes the distinction between the classes `0' or `1' and  `1' or `2' may seem arbitrary. This, however, does not affect the subsequent discussion.

The third column of the `Remarks' refers to the presence of light-curve variation for fundamental-mode RRab stars (see details in Paper II). `Bl' denotes that the star definitely exhibits Blazhko effect, while `Bl:' indicates that light-curve modulation is probable. RRab stars whose photometry is seriously affected by crowding or close, bright, neighbouring star(s) are labelled by `d'. The deficiency of the photometry of these stars makes any statement on their light-curve variation impossible.

The last four columns of Table~\ref{tabla} give information on the observed period changes of the cluster's RR Lyrae stars: $\beta = {{\mathrm{d}}P/{\mathrm{d}}t}$ in $10^{-10}$  and dMyr$^{-1}$ and $\alpha = P^{-1} {{\mathrm{d}}P/{\mathrm{d}}t}$ in ${\mathrm{10^{-10}d^{-1}}}$  units, while $P_{\mathrm{max}}-P_{\mathrm{min}}$ is the full range (the maximum fluctuation) of the observed period variation.

We have to admit that in some cases it is difficult to decide whether the period is constant with some random noise or it has a very slow monotonic change. If the absolute value of the coefficient of the quadratic term in the parabolic approximation was smaller than its $2\sigma$ error then the $O-C$ diagram was fitted by a straight line, and $\beta = 0$ was accepted. Of course, for these stars the formally calculated full range of the period variation may differ slightly from zero due to small scatter originating from real period noise or observational error.

\subsection{Evolutionary considerations}

The final goal of this study is to determine if the observed period changes of the M5 RR Lyrae stars could be attributed, at least in the mean, to stellar evolution. From evolutionary consideration it is well-known that the period increases as a star evolves redward, and decreases as it moves blueward in the HR diagram. Theory predicts evolution in both directions and with different period-change rates depending on the HB-evolutionary stage of the RR Lyrae stars. It has also been known for some time that RR Lyrae period changes may be characterized by strong irregularities. As Table~\ref{tabla} shows, more than one third of the $O-C$ diagrams of M5 RR Lyrae stars cannot be interpreted solely by constant or linearly changing periods. In spite of the fact that evolutionary changes are masked by irregularities in a number of cases, it is hoped that the period-change rates of RR Lyrae stars with steady period variation give some information on their evolution.

\begin{table}
\caption{Statistics of the pulsation-period-change rates of RR Lyrae stars in M5}
 \label{beta}
  \begin{tabular}{llllc}
\hline
& &$ \beta \,  [\mathrm{dMyr^{-1}}]$ & $\alpha \, [\mathrm{Myr^{-1}}]$& No.\\
\hline
\multicolumn{5}{l}{total sample}\\
\hline
RRc& mean        & $-0.003\pm0.073$ &$-0.014\pm0.221$&11\\
   & m.p.v.$^{a}$& $-0.006$         &$-0.021$       &  \\
RRab&mean        & $-0.006\pm0.178$ &$-0.023\pm0.328$&44\\
    &m.p.v.      & $+0.003$         &$+0.001$       &  \\
\hline
all&mean         & $-0.006\pm0.162$ &$-0.021\pm0.308$&55\\
   &m.p.v        & $+0.002$         &$-0.002$       &  \\
\hline
\hline
\multicolumn{5}{l}{without 3 outliers}\\
\hline
RRab&mean        & $+0.004\pm0.123$ &$-0.005\pm0.234$&41\\
    &m.p.v.      & $+0.008$         &$+0.012$       &  \\
\hline
all&mean         & $+0.003\pm0.114$ &$-0.007\pm0.229$&52\\
   &m.p.v        & $+0.005$         &$+0.005$       &  \\
\hline
\hline
\multicolumn{5}{l}{without Blazhko RRab stars}\\
\hline
RRab&mean        & $+0.017\pm0.163$ &$+0.021\pm0.296$&38\\
    &m.p.v.      & $+0.015$         &$+0.024$       &  \\
\hline
all&mean         & $+0.012\pm0.147$ &$+0.013\pm0.279$&49\\
   &m.p.v        & $+0.009$         &$+0.014$       &  \\
\hline
\hline
\multicolumn{5}{l}{\footnotesize{$^{a}$ the most-probable value, i.e. the mode}}\\
\end{tabular}
\end{table}

\begin{figure}
\includegraphics[angle=0,width=9cm]{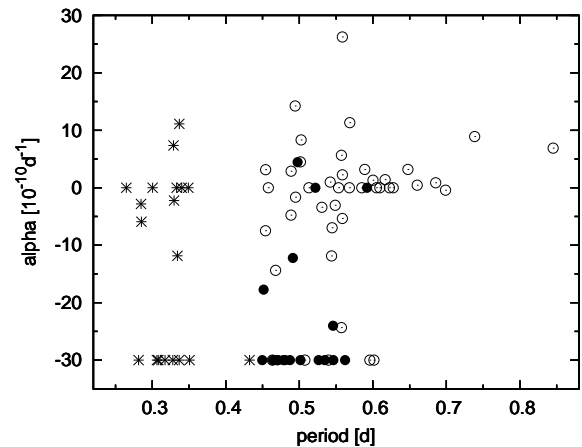}
\caption{Period-change rates ( $ { \alpha = \dot{P} /P {\mathrm{[10^{-10}d^{-1}] }}} $ ) 
of the RR Lyrae stars in M5. Circles and asterisks denote fundamental-mode and first-overtone variables, respectively. RRab stars showing the Blazhko effect are shown by filled symbols. Variables with irregular period-change behaviour are set at $\alpha = -30$ to show their period distribution. Positive and negative period-change rates are equally frequent among M5 variables, with extreme ($|\alpha| > 20$) values for three RRab stars. Period decreases, irregular period change and Blazhko modulation occur in RRab stars only with period shorter than 0.6 d. }
\label{figalfa}
\end{figure}

\begin{figure}
\includegraphics[angle=0,width=9cm]{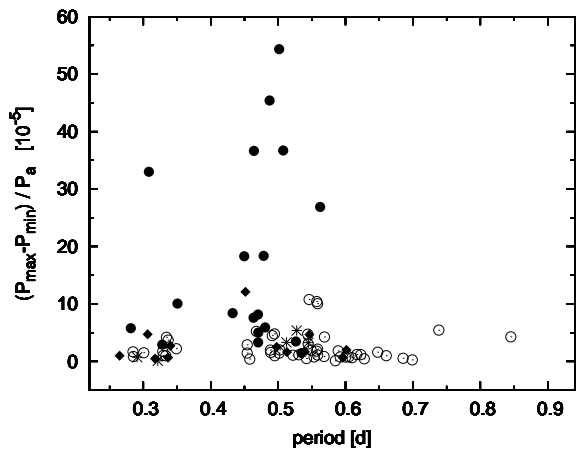}
\caption{Maximum values of the normalised period variation of the RR Lyrae stars in M5. Filled circles and diamonds denote variables with strong and moderate irregularity of their period changes, stars with regular period variations (constant or linear) are shown by open circles, asterisks are for variables with not enough information on the period-change properties. Period changes of the order of  $10^{-4}$~d are observed in both RRab and RRc stars. These very large values of period changes are detected only in variables with irregular period changes and in RRab star only at shorter periods.
}
\label{deltap}
\end{figure}

The mean and the most-probable values of the period-change rates for the RRc, RRab and the total sample of RR Lyrae stars that have straight-line or parabolic $O-C$ diagrams are summarized in the first part of Table~\ref{beta}. Although the averages of the $\alpha$ and $\beta$ values suggest a slight, overall period decrease, it is not the least significant. The most-probable values of the period changes of the statistically larger samples (RRab and all the variables) are practically zero.

The generally accepted HB-evolutionary models do not indicate that high, negative period-change rates are likely, and large, positive values are supposed to occur only in the very late stages of HB evolution at luminosities brighter than on the zero-age horizontal branch (ZAHB). The period-change rate $\alpha$ versus period for the M5 variables is shown in  Fig.~\ref{figalfa}, and the largest value of the normalised period variation versus period is plotted in Fig.~\ref{deltap}. From these figures it can be immediately seen that in M5 (like in other globular clusters) there are variables with extreme $\alpha$ values. These extreme period changes occur among RRab rather than among RRc stars according to the study of \cite{rs}.

In our sample, there are three RRab stars with extreme period-change rates, two stars (V5 and V81) with large negative and one (V70) with large positive values. V70 lies far from the cluster centre, so it was not included in any of the CCD studies. Therefore, only its photographic $B$ magnitude is available. The mean $B_{pg}$ brightness of V70 is one of the faintest among the whole sample, indicating that V70 cannot be in a highly evolved evolutionary phase. We thus suppose that the period changes of these three stars are not induced by evolutionary effects. Since their extreme values strongly influence the mean period-change rates, the averages and the most-probable values of the period changes are also derived excluding them (see the second part of Table~\ref{beta}). The most-probable $\alpha$ and $\beta$ values for RRab and for all the 52 stars correspond to small positive rates, while the mean values  have very small positive and negative $\beta$  and $\alpha$ values, respectively.

We thus conclude that, on the average, the period-change rate of RR Lyrae stars in M5 are very close to zero. We also note that the median rates of the period changes for all the different samples is zero. This conclusion confirms the result of the previous studies \citep{cs69,re96} on an extended baseline and larger sample, and  is in good agreement with the prediction of Lee's (1991) calculation on the stellar evolution -- period change connection: the mean (median) rate of period change should be close to zero (0.005--0.010 dMyr$^{-1}$) for clusters with HB-type like M5.

\begin{figure}
\includegraphics[angle=0,width=8.9cm]{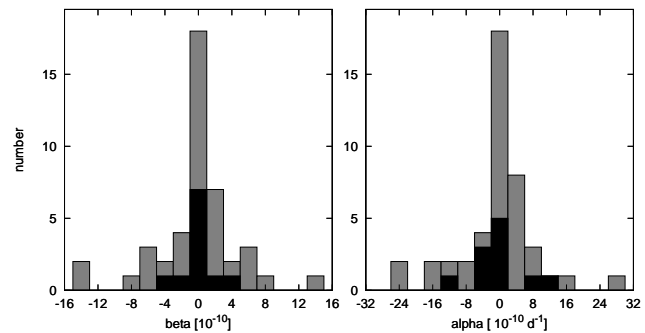}
\caption{Histograms of the distributions of the period-change rates  $\beta$ [$\mathrm{10^{-10}}$] and $\alpha$ [$\mathrm{10^{-10} d^{-1}}$]. RRab and RRc stars are shown by gray and black areas, respectively. Although there is some hint that small positive period values are in excess to negative ones in RRab stars, this effect statistically is only weakly significant. On the average, the period-change rates of M5 variables are very close to zero. }
\label{hist}
\end{figure}

An other aspect of the problem of period changes is the distribution of the direction of the changes, whether the proportion of the decreasing and increasing periods reflects the evolutionary state of the RR Lyrae stars of a globular cluster. The models of post-ZAHB evolution \citep{ldz,dor} predict that after an initial `hook' to the red caused by shell readjustment, RR Lyrae stars evolve blueward (with decreasing periods) across the instability strip, and that when they have depleted the helium in their cores, they increase in luminosity and evolve redward (with increasing periods). The histograms of the distribution of the period-change rates $\beta$ and $\alpha$ are shown in Fig.~\ref{hist}. This figure suggests a slight excess of RRab stars with small increasing period. Our sample has 19 ab stars with increasing period, contrasting with 14 ab variables with decreasing period, while 11 ab stars have constant period. Among the RRc stars 2 have increasing, 4 decreasing and 5 constant period. Hence, the full sample contains 21 variables with increasing, 18 with decreasing and 16 with constant period. The small number of the c-type stars obviously precludes the possibility of their distinct statistical treatment.

A simple test can prove that the observed small excess of RRab stars with increasing period is only weakly significant. If we distribute equally the stars with $\beta=0$, we end up with 24 or 25 positive and 20 or 19 negative. Supposing that the likelihood of the positive and negative period change is equal, the expected value of the binomial distribution is 22 with a variance of 3.32. For the whole sample, the expected value is 27.5 with a variance of 3.71, so the distribution of the $\beta$ value signs does not provide clear statistical evidence for an excess of period increases among all the M5 RR Lyrae stars. This might be a consequence of the fact that the mean period-change rate of the variables is in agreement with the $\approx$zero mean period-change rates of the variables. 

In M3, another OoI-type cluster, \cite{cc01} found 21 RRab stars with increasing and 14 with decreasing period, i.e. a small excess of increasing periods. This might be explained by that a larger fraction of the RRab stars of M3 shows OoII characteristics \citep{m3} than that of the M5 RRab variables. At the same time, the RR Lyrae stars of the well-observed OoII-type clusters like M15 \citep{ss95} and $\omega$ Cen \citep{ocen} show a definite surplus of increasing periods. This implies that the majority of the variables in these clusters are evolving across the instability strip from blue to red. All these results strengthen that the Oosterhoff properties of globular-cluster variables are strongly connected to their horizontal-branch evolutionary state \citep{cs99}. Nevertheless, it should be noted that \cite{ss81} called attention to the fact that no cluster exhibits a large plurality of decreasing period.

On the whole, the HB evolution is a plausible explanation for increasing periods. Stars within the instability strip evolve rapidly from blue to red, toward the end of core helium burning, producing a large positive rate of period change. In this respect, it is worth mentioning the difference between the period-change behaviour of the RRab stars of the clusters M3 and M5. Although both clusters belong to Oosterhoff-type I and resemble each other in many aspects, M5 has fewer strongly increasing periods than M3. In M5, only five out of the 44 RRab (11 per cent) have very large period-increase rate ($\beta>0.20$ dMyr$^{-1}$), while in M3, 29 per cent of the RRab stars have. In other OoI-type clusters, M14 \citep{wf94}, M28 \citep{w86} and NGC 7006 \citep{w99} the frequency of RRab stars with strongly increasing periods are 16, 8 and 10 per cent, respectively. This simple statistic shows that the M5 RRab stars behave like other OoI-type clusters, and only M3 behaves somewhat differently.

\begin{figure}
\includegraphics[angle=0,width=9cm]{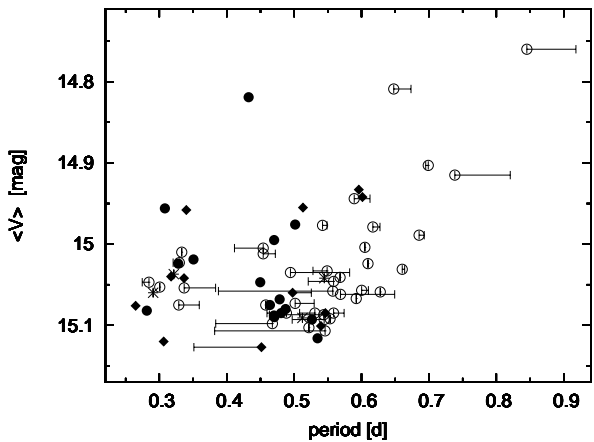}
\caption{Intensity-averaged mean $V$ brightness versus period diagram of the RR Lyrae stars in M5. Symbols are the same as in Fig.~\ref{deltap}. The period-change rate and its direction are indicated by horizontal lines. RRab stars with period decrease are characterized by short period and fainter luminosities, large period increase occurs at longer period, more luminous variables.  }
\label{vp}
\end{figure}

\begin{figure*}
\includegraphics[angle=0,width=18.5cm]{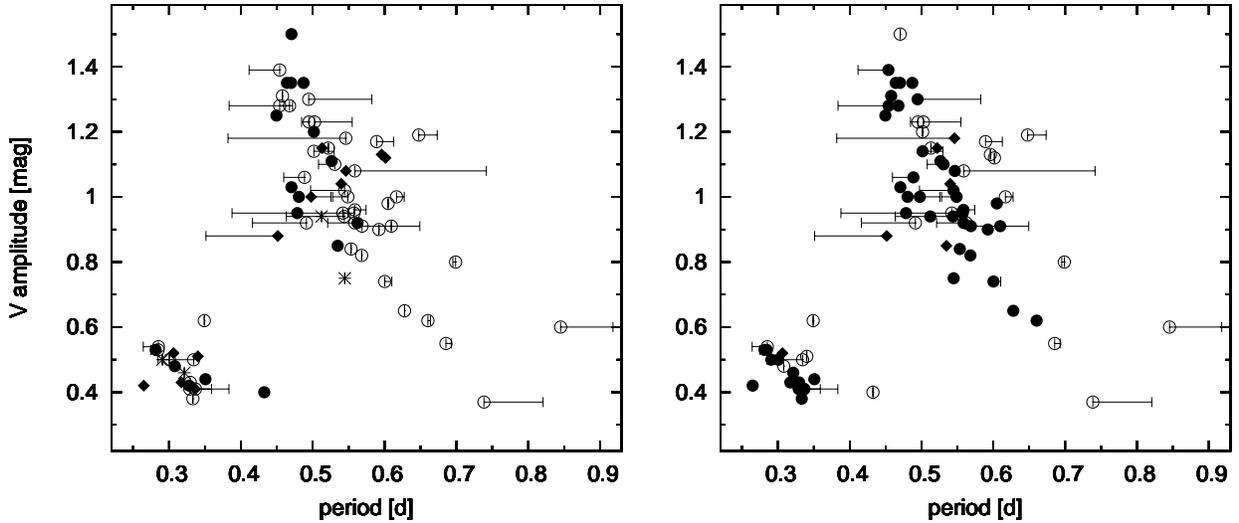}
\caption{$V$ amplitude versus period plots of the RR Lyrae stars in M5. In the left-hand panel symbols indicate the regular and irregular period-change behaviour of the stars as in Fig.~\ref{vp}. In the right-hand panel the symbols indicate the intensity-averaged mean $V$ brightness of the variables, open circle: $\langle V\rangle < 15\fm0$; filled circle: $15\fm0 < \langle V\rangle <  15\fm1$; filled diamond: $\langle V\rangle > 15\fm1$. The period-change rate and its direction is shown by horizontal lines in both panels. In the left-hand panel we can also recognize that the period changes of short-period, large-amplitude RRab stars are often irregular. Large negative period changes occur in short-period RRab stars at fainter mean magnitudes while large period increases are observed at the whole period range of RRab stars, but the amplitudes of these variables tend to be larger than that of the majority of the stars, indicating their evolved HB stage. }
\label{vap}
\end{figure*}

One of the most intriguing issues connected with the period changes of RR Lyrae stars is the large negative values of $\beta$, observed in different clusters. Among the 44 RRab stars of M5 that have parabolic or straight-line $O-C$ diagrams, nine (20 per cent) show decreasing periods with as high rate as $\beta<-0.1$ dMyr$^{-1}$, that certainly cannot be explained by canonical HB evolution. It is worth mentioning that the mean period of the sample of 44 RRab stars is 0.5622 d, whereas the mean period of the nine ab stars with fast decreasing period is significantly less, 0.5127 d.

A comparison with other well-studied globular clusters proves that the occurrence of RRab stars with very fast period decreases is a general feature, characteristic of both Oosterhoff-types. For example, 11 out of the 35 RRab stars (31 per cent) in M3 \citep{cc01} and 9 out of the 42 ab variables (21 per cent) in NGC 6934 \citep{sw} exhibit large period decreases ($\beta<-0.10$ dMyr$^{-1}$). The OoII-type clusters also have (perhaps relatively less in numbers) RRab stars with strong period decreases. In $\omega$ Cen, 29 RRab stars of the chemically homogeneous group have linear period changes, and among them, four (14 per cent) have strong period decreases \citep[table 7 in][]{ocen}. In M15, there are two out of 13 RRab stars (15 per cent) that show strong period decreases \citep{ss95}. The trend that the mean period of the ab stars with fast period decreases is shorter than the mean period of the whole sample also holds for the clusters mentioned. 

To resolve the contradition between canonical HB evolutionary model predictions and the observed frequency of strong period decreases, it is generally supposed that most of the period changes are due to some kind of `noise' rather than to evolutionary effects. \cite{sr79} have found that the mixing events at the convective core edge, the transfer of helium into the convective core and the subsequent chemical readjustment of the semi-convective zone around the core, is an intrinsically noisy process that can lead to large period-decrease rates. 

Recently, \cite{si08} have given an alternative explanation for the frequent occurrence of strong period decreases among the cluster variables. They constructed pre-ZAHB evolutionary tracks for a chemical composition appropriate to the globular cluster M3, and investigated the period-change behaviour of variables in the final approach to the ZAHB location. They have found that, before settling on the ZAHB, the variables are subject to a strong period decrease with the most likely $\beta$ values around $-0.3$ dMyr$^{-1}$, but more extreme values ($-0.8$ dMyr$^{-1}$) may also take place. The model simulations have also shown that some percent of the RR Lyrae population are in the pre-ZAHB evolutionary state, and the pre-ZAHB pulsators are expected to have longer periods than the bona fide HB pulsators. As M5 resembles M3 in different aspects, therefore the model calculations of \cite{si08} can be compared with our results. The high percentage of variables with decreasing periods and the relatively shorter mean periods of these stars in M5 show that, likely, the pre-ZAHB evolution cannot fully explain the observations. Nevertheless, further model calculations of the pre-ZAHB evolutionary phase may lead to more satisfactory results.

It is perplexing, however, why the observed period-increase rates are in very good agreement with evolutionary predictions, if the period-decrease rates are not. If this is indeed the case, some significant difference between RR Lyrae stars with increasing and decreasing period rates should be found, but observationally, there is no evidence of any difference between these stars.

It is an interesting question whether the evolutionary effects are apparent in the different diagrams of cluster variables, e.g. period-amplitude, period-brightness, etc. diagrams. Fig.~\ref{vp} and Fig.~\ref{vap} show the relationships between the intensity-averaged mean $V$ brightness $\langle V \rangle$ and period, as well as between the $V$ amplitude ($A_V$) and the period. The period-change rate and its direction are indicated by horizontal lines in both figures. The $\langle V \rangle$ versus period plot displays that the variables with longer periods have higher luminosity and constant or increasing period (the only exception is V9 with a very slow $\beta = -0.011$ dMyr$^{-1}$ decrease). The long-period sequence in the $A_V$ versus period diagram shows the same trend. In accordance with evolutionary theories these stars are at an advanced evolutionary stage moving off from the HB \citep{cs99}.

One of the most important issues that should be addressed, is the irregular period changes of RR Lyrae stars in globular clusters. Scrutinizing the $O-C$ diagrams of 81 RR Lyrae stars in M5, it turned out that 31 out of them (38 per cent) cannot be satisfactorily fitted by straight line or parabola. According to the number of inflection points and the order of the polynomial fit, these diagrams were classified as moderately or strongly irregular (see index `1' and `2' in the middle column of the `Remarks' in Table~\ref{tabla}). If we separately investigate the irregular period behaviour of the fundamental-mode and first-overtone pulsators, we find that more than the half of the c-type stars (10/18) have complex $O-C$ diagrams, while one third of the RRab stars (21/63) have irregular period changes. The fact that irregular period behaviour is more frequent among RRc stars than RRab stars has already been noted by \cite{ocen} in a similar study of the period-change behaviour of the $\omega$ Cen variables. A comparison with other clusters' RRab and RRc stars prove that this period behaviour tends to be a general feature. In M14 \citep{wf94} and in NGC 7006 \citep{w99} the frequency of irregular period changes of ab-stars are 20 and 14 percentages, while those of the c-type stars are 80 and 43 percentages, respectively.

The most generally accepted explanation for the irregular period changes was put forward by \cite{sr79}. The mixing process within the semi-convective zone and the overshooting at the convective-core edge were found to be responsible for the non-evolutionary changes. These physical processes, however, cannot account for the different period-change behaviour of the RRab and RRc stars. Another problem is that the observed irregular period changes are as large as 2--5\,$\cdot10^{-4}$ ${\Delta P} {P^{-1}}$ in some cases (e.g. in V4, V14, V18, V52 and V72) occurring on time-scales of some tens of years only. Although many parameters influence the time-scales and the resultant period-change values caused by mixing events, as an average value, 370 yr was derived for the time interval required to produce a period change of $3\cdot10^{-5}$ (${\Delta P} {P^{-1}}$) \citep[see fig. 9 in][]{sr79}. The two orders of magnitude discrepancy between observations and model predictions may give rise to some doubt that the rapid, irregular period variations of RR Lyrae stars can indeed be generated by the proposed mixing events in the interior. To settle this issue, more detailed numerical modelling of the effect of the mixing events on the pulsation period is highly needed.

\begin{figure}
\includegraphics[angle=0,width=9cm]{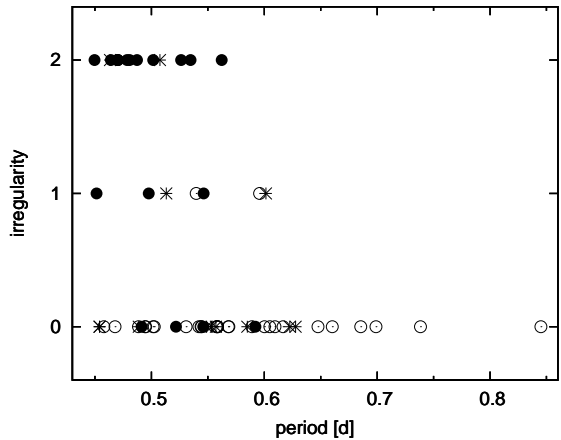}\caption{Connection of irregular period changes with Blazhko effect of RRab stars. Stars with strong and moderate irregular period changes are set at 2 and 1 values. Stars with constant period and with linear period changes are plotted at 0. Filled and open circles denote Blazhko and mono-periodic RRab stars. Variables, which have scarce and defective photometry are denoted by asterisks. Light-curve modulation of these stars cannot be detected because of the inaccuracy of their photometry. There is a striking coincidence of variables with complex period-change behaviour and with the occurrence of light-curve modulation.} 
\label{irreg}
\end{figure}

Here we outline another approach to the interpretation of irregular period changes of RR Lyrae stars. Fig.~\ref{irreg} shows that there is a striking connection between the irregular period changes and the Blazhko effect of M5 RR Lyrae stars. We omit the RRc stars from the following discussion, the detection of Blazhko effect in these stars is difficult as the modulation amplitude is usually in the range of the photometric accuracy.  The figure demonstrates that all RRab stars exhibiting very complex $O-C$ diagram, i.e. irregular period changes, have been found to show the Blazhko effect, if the sample of variables with good enough photometry to detect light-curve variability is considered.
 
It seems that the inverse statement is also true with only a few exceptions. Two Blazhko stars, V1 and V30 have constant period, and two, V5 and V58 have decreasing period with very high rates, $-0.479$ dMyr$^{-1}$ and $-0.219$ dMyr$^{-1}$, respectively. If, however, the $O-C$ diagrams of field Blazhko RR Lyrae stars are considered, it can be immediately realized that the irregular (from time to time abrupt) period changes are followed by a long interval of linearly changing period, a period readjustment \cite[see e.g. the $O-C$ diagrams of XZ Cyg, RR Gem, SZ Hya in fig. 4 of][]{geos}. On the basis of all this, the period changes of Blazhko RR Lyrae stars cannot be identified with evolutionary effects, and should be left out of evolutionary considerations.

If variables showing Blazhko effect are omitted then the period-change rates of RRab stars with constant or linearly changing period (38 stars) have small positive mean and most-probable values. If RRc stars are added to this sample (49 stars) the mean and the most-probable values of the period-change rates hardly change. The last section of Table~\ref{beta} summarizes the mean period-change rates and the most-probable values if the Blazhko RRab stars are left out of consideration. Curiously enough, these values are in excellent agreement with HB-model predictions of \cite{le91}.

The distribution of the period-change rates of the RRab stars also changes when the Blazhko stars are omitted. Now, among the 38 RRab stars 11, 9 and 18 have decreasing, constant and increasing period-change rates, respectively. In this case, a binomial distribution with equal likelihood of both period-change directions has a variance of 3.1, while the difference between period changes is 7, more than twice as large. So, in this sample, the period of RRab stars is dominantly increasing.

\section{Summary}
The period changes of eighty-six M5 RR Lyrae stars have been investigated. The published data have been supplemented by archival, previously unpublished observations covering the second half of the last century. The observations of M5 variables now provide an almost continuous time coverage for about a hundred-year time base, making the accurate study of the period changes possible.

Reliable $O-C$ diagrams (phase diagrams) were constructed for 81 RR Lyrae, 62 RRab and 19 RRc stars. The $O-C$ plots of 44 RRab and 11 RRc stars could be well approximated by a straight line or parabola, indicating constancy or linear change of the pulsation period of these stars on the hundred-year time base of the observations. 21 RR Lyrae have increasing, 18 decreasing and 16 constant period. The excess of increasing period among RRab stars is slightly more pronounced, however, statistically that is still weakly significant. Omitting Blazhko variables the situation changes, there are 18 stars with increasing, 9 with constant and only 11 with decreasing period in this sample.

The average, median and most-probable values of the period-change rates of M5 variables equals zero, or have very small, either positive or negative values for different subsamples. Omitting the Blazhko stars, however, the period-change rates indicate a small period increase on the average. Its  most-probable value for 49 stars is 0.009 dMyr$^{-1}$, in very good accordance with predictions of synthetic HB-model results \citep{le91}.

Large period-change rates (both negative and positive) characterize some of the shorter-period RRab stars, while variables with period longer than 0.6 d always have constant or increasing period. The observed rates of period increase are, in general, in good agreement with standard HB evolutionary model predictions, but most of the period-decrease rates are significantly larger than that canonical models would allow. 
The recent supposition of \cite{si08} that connects these changes to pre-ZAHB evolution is a promising attempt,  but fails to explain the statistics and the period distribution of these stars.

More than one third of the $O-C$ diagrams of M5 RR Lyrae stars cannot be fitted solely by straight line or parabola, i.e. the period changes of these stars are moderately or strongly irregular. The pulsation periods of some of these stars vary by some $10^{-4}$ d within relatively short time intervals (10--30 yr). The irregular behaviour of the periods is more frequent among RRc stars, and it is rather characteristic of the shorter-period RRab stars. An important relation between the irregular period change and the Blazhko effect has been revealed: in M5, stars with strongly irregular period changes always have Blazhko modulation.

\section*{Acknowledgments}
The constructive, helpful comments of the referee, Katrien Kolenberg are much appreciated. We thank Dr. Katalin Ol\'ah for placing her photographic observations at our disposal. The financial support of OTKA grant K-068626 is acknowledged. C. Clement thanks the Natural Sciences and Engineering Research Council of Canada for financial support. Zs. H. thanks the `Lend\"ulet' program of the Hungarian Academy of Sciences for supporting his work.


\begin{thebibliography}{99}
\bibitem[Arp(1955)]{ar55} Arp, H., 1955, AJ, 60, 317
\bibitem[Arp(1962)]{ar62} Arp, H., 1962, ApJ, 135, 311
\bibitem[Bailey(1917)]{ba17} Bailey, S.I., 1917, Harvard Ann., 78, 103
\bibitem[Bal\'azs-Detre \& Detre(1965)]{bd65} Bal\'azs-Detre, J., Detre, L., 1965, in Strohmeier, W. ed, The Position of Variable Stars in the Hertzsprung-Russell diagram, Proc. IAU Coll. No.3, Ver\"off. der Remeis-Sternwarte Bamberg IV. (Nr. 40), 184
\bibitem[Brocato et al.(1996)]{br96} Brocato, E., Castellani, V., Ripepi, V., 1996, AJ, 111, 809
\bibitem[Clement et al.(1992)]{cj92} Clement, C., Jankulak, M., Simon, N. R., 1992, ApJ, 395, 192
\bibitem[Clement \& Shelton(1999)]{cs99} Clement, C., Shelton, I., 1999, ApJ, 515, L85
\bibitem[Cohen \& Gordon(1987)]{cg87} Cohen, J.G., Gordon, G.A., 1987, ApJ, 318, 215 
\bibitem[Cohen \& Matthews(1992)]{cm92} Cohen, J.G., Matthews, K., 1992, PASP, 104, 1205
\bibitem[Corwin \& Carney(2001)]{cc01} Corwin, T. M., Carney B. W., 2001, AJ, 122, 3138
\bibitem[Coutts(1971a)]{co71a} Coutts, Ch. M., 1971a, in Strohmeier, W. ed, New Directions and New Frontiers in Variable Star Research, Proc. IAU Coll. No.15, Ver\"off. der Remeis-Sternwarte Bamberg IX.(Nr.100.), 238
\bibitem[Coutts(1971b)]{co71b} Coutts, Ch. M., 1971b, Publ. David Dunlap Observatory 3(3), 79 
\bibitem[Coutts \& Sawyer Hogg(1969)]{cs69} Coutts, Ch. M., Sawyer Hogg, H., 1969, Publ. David Dunlap Observatory 3(1), 1 
\bibitem[Cuffey(1956)]{c56} Cuffey, J., 1956, Sky and Telescope, 15, 258
\bibitem[Cuffey(1961)]{c61} Cuffey, J., 1961, AJ, 66, 71
\bibitem[Dorman(1992)]{dor} Dorman, B., 1992, ApJS, 81, 221
\bibitem[Eddington(1918)]{ed18} Eddington, A. S., 1918, MNRAS, 79, 2
\bibitem[Jurcsik et al.(2005)]{ibvs} Jurcsik, J., S\'odor, \'A., V\'aradi, M., 2005, IBVS, No.5666
\bibitem[Jurcsik et al.(2001)]{ocen}  Jurcsik, J. et al., 2001, AJ, 121, 951
\bibitem[Jurcsik et al.(2003)]{m3} Jurcsik, J. et al., 2003, ApJ, L597
\bibitem[Jurcsik et al.(2010)]{p2} Jurcsik, J. et al., 2010, MNRAS, submitted (Paper II)
\bibitem[Kaluzny et al.(2000)]{ka00} Kaluzny, J. et al., 2000, A\&AS, 143, 215 
\bibitem[Koopmann et al.(1994)]{k94} Koopmann, R., Lee, Y.-W., Demarque, P., Howard, J. M., 1994, ApJ, 423, 380
\bibitem[Kukarkin \& Kukarkina(1971)]{kk71} Kukarkin, B.V., Kukarkina, N.P., 1971, Perem. Zvezdy Pril., 1(No.1), 1 
\bibitem[Laskarides(1974)]{la74} Laskarides, P. G., 1974, Ap\&SS, 27, 485
\bibitem[Le Borgne et al.(2007)]{geos} Le Borgne, J.-F. L.  et al., 2007, A\&A, 476, 307
\bibitem[Lee(1991)]{le91} Lee, Y.-W., 1991, ApJ, 367, 524
\bibitem[Lee \& Demarque(1990)]{ld} Lee, Y.-W., Demarque, P., 1990, ApJS, 73, 709
\bibitem[Lee et al.(1990)]{ldz} Lee, Y.-W., Demarque, P., Zinn, R., 1990, ApJ, 350, 155
\bibitem[Lombard \& Koen(1993)]{lk93} Lombard, F., Koen, Ch., 1993, MNRAS, 263, 309
\bibitem[Martin(1938)]{ma38} Martin, W. Chr., 1938, Leiden Ann., 17, No.1
\bibitem[Oosterhoff(1941)]{oo41} Oosterhoff, P. Th., 1941, Leiden Ann., 17, No.4 
\bibitem[Rathbun \& Smith(1997)]{rs} Rathbun, P. G., Smith, H. A., 1997, PASP, 109, 1128
\bibitem[Reid(1996)]{re96} Reid, I.N., 1996, MNRAS, 278, 367, Erratum: 1996 MNRAS, 282, 304 
\bibitem[Shapley(1927)]{sh27} Shapley, H., 1927, Harvard Bull., 851, 15 
\bibitem[Silbermann \& Smith(1995)]{ss95} Silbermann, N. A., Smith, H. A., 1995, AJ, 109, 1119
\bibitem[Silva Aguirre et al.(2008)]{si08} Silva Aguirre, V. et al., 2008, A\&A, 489, 1201
\bibitem[Smith \& Sandage(1981)]{ss81} Smith, H. A., Sandage, A., 1981, AJ, 86, 1870
\bibitem[Stagg \& Wehlau(1980)]{sw} Stagg, Ch., Wehlau, A., 1980, AJ, 85, 1182
\bibitem[Sterne(1934)]{st34} Sterne, Th.E., 1934, Harvard Obs. Circ., No.386-387
\bibitem[Storm et al.(1991)]{st91} Storm, J., Carney, B.W., Beck, J.A., 1991, PASP, 103, 1264 
\bibitem[Stothers(1980)]{st80} Stothers, R., 1980, PASP, 92, 475
\bibitem[Sweigart \& Renzini(1979)]{sr79} Sweigart, A.V., Renzini, A., 1979, A\&A, 71, 66
\bibitem[Wehlau et al.(1986)]{w86} Wehlau, A. et al.,  1986, AJ, 91, 1340
\bibitem[Wehlau \& Froelich(1994)]{wf94} Wehlau, A., Froelich, N., 1994, AJ, 108, 134
\bibitem[Wehlau et al.(1999)]{w99} Wehlau, A., Slawson, R.W., Nemec, J. M.,  1999, AJ, 108, 134
\end{thebibliography}
\end{document}